\documentclass[10pt,twocolumn,article]{IEEEtran}
\usepackage{times}
\usepackage{amsbsy}
\usepackage{latexsym}
\usepackage{amssymb}
\usepackage{mathtools}
\usepackage{amsmath,amssymb,amsfonts,graphicx,epsfig,epstopdf}
\usepackage{times,amsbsy,latexsym,bm,color}
\usepackage[ansinew]{inputenc}

\title{Direction Finding Based on Multi-Step Knowledge-Aided Iterative Conjugate Gradient
Algorithms}

\author{Silvio F. B. Pinto $^1$ and Rodrigo C. de Lamare $^{1,2}$ \\
Center for Telecommunications Studies (CETUC) \\ $^1$ Pontifical
Catholic
University of Rio de Janeiro, RJ, Brazil.\\
$^2$ Department of Electronics, University of York, UK \\
Emails: silviof@cetuc.puc-rio.br, delamare@cetuc.puc-rio.br }
\begin{document}
\maketitle
\begin{abstract}
In this work, we present direction-of-arrival (DoA) estimation
algorithms based on the Krylov subspace that effectively exploit
prior knowledge of the signals that impinge on a sensor array. The
proposed multi-step knowledge-aided iterative conjugate gradient
(CG) (MS-KAI-CG) algorithms perform subtraction of the unwanted
terms found in the estimated covariance matrix of the sensor data.
Furthermore, we develop a version of MS-KAI-CG equipped with
forward-backward averaging, called MS-KAI-CG-FB, which is
appropriate for scenarios with correlated signals. Unlike current
knowledge-aided methods, which take advantage of known DoAs to
enhance the estimation of the covariance matrix of the input data,
the MS-KAI-CG algorithms take advantage of the knowledge of the
structure of the forward-backward smoothed covariance matrix and its
disturbance terms. Simulations with both uncorrelated and correlated
signals show that the MS-KAI-CG algorithms outperform existing
techniques.
\end{abstract}

\section{Introduction}
\label{introduction}

In sensor array signal processing, direction-of-arrival (DoA)
estimation is a topic of fundamental importance for applications in
wireless communications, radar and sonar systems, and seismology
\cite{Vantrees1,locsme,elnashar,manikas,cgbf,okspme,r19,scharf,bar-ness,pados99,
reed98,hua,goldstein,santos,qian,delamarespl07,delamaretsp,xutsa,xu&liu,
kwak,delamareccm,delamareelb,wcccm,delamarecl,delamaresp,delamaretvt,delamaretvt10,delamaretvt2011ST,
delamare_ccmmswf,jidf_echo,jidf,barc,lei09,delamare10,fa10,ccmavf,lei10,jio_ccm,
ccmavf,stap_jio,zhaocheng,zhaocheng2,arh_eusipco,arh_taes,rdrab,dcg,dce,dta_ls,
song,wljio,barc,saalt,mmimo,wence,spa,mbdf,rrmber,bfidd,did,mbthp,wlbd,baplnc}.
Existing high-resolution algorithms for DoA estimation such as the
multiple signal classification (MUSIC) method \cite{schimdt}, the
root-MUSIC algorithm \cite{Barabell}, the estimation of signal
parameters via rotational invariance techniques (ESPRIT) \cite{Roy}
and other recent subspace techniques \cite{Steinwandt,Wang,Qiu} rely
on the estimation of the signal and orthogonal subspaces from
estimates of the covariance matrix of the sensor data. {After years
using these and other methods for DoA estimation and their versions,
all of them based on the estimate of the covariance matrix, two
emerging fields of research have attracted interest of academics.
The former is the employment of sparse configurations like nested
\cite{Pal1} and coprime \cite{Pal2} arrays in order to obtain
$\mathcal{O}\left( \mathit{M^{2}}\right) $ degrees-of- freedom
(DOFs) with only $ \mathit{M}$ physical sensors, which can be viewed
as sensors efficiency in direction finding. Recent works based on
compressive sensing \cite{Gu1} and sparse reconstruction\cite{Gu2}
also applied to coprime arrays represent current efforts on this
topic.  The latter is a promising emerging field of research, which
is related to the present work and focuses on the accuracy of the
covariance matrix of the sensor data and its impact on high
resolution DoA estimation.}

Prior work on conjugate gradient (CG) techniques include the works
\cite{cgbf,smcg,rdrcb,Semira,Steinwandt}. Early work with CG
algorithms include element-based direction finding \cite{Semira} and
beamspace approaches \cite{Steinwandt}, respectively. Previously
reported work on knowledge-aided techniques include the Two-Step
Knowledge-Aided Iterative ESPRIT (TS-KAI-ESPRIT) \cite{Pinto} and
the Multi-Step Knowledge-Aided ESPRIT (MS-KAI-ESPRIT)
{\cite{Pinto1,Pinto4}}, which enhance the covariance matrix
estimates using subtraction of the unwanted terms \cite{Vorobyov1,
Vorobyov2}. In particular, these approaches compute a subtraction
factor that mitigates the effect of the unwanted terms on the
estimation of the covariance matrix. As a result of the subtraction
of the unwanted terms the estimates of the covariance matrix are
enhanced. This subtraction procedure is performed by using the
likelihood function and choosing the set of DoA estimates that
maximize it. TS-KAI-ESPRIT performs this subtraction using only two
steps and requires perfect prior knowledge of signals whereas
MS-KAI-ESPRIT employs the same number of steps as the number of
signals\cite{Pinto2,Pinto3,Steinwandt2,Stoica2}. Specifically, the
prior knowledge used in previously reported works often rely on
statistical quantities such as the covariance matrix of the signals
arising from known static users in a system. In contrast,
MS-KAI-ESPRIT acquires its prior knowledge on line, i.e., by means
of preliminary estimates, computed at the first step. At each
iteration of its second step, the initial steering matrix is updated
by replacing a growing number of steering vectors of initial
estimates for their corresponding newer ones. That is to say, at
each iteration, the additional knowledge acquired on line is
updated, allowing MS-KAI-ESPRIT to introduce updates in the sample
covariance matrix estimate, which results in improved estimates.

In this paper, we present the Multi-Step Knowledge-Aided Iterative
Conjugate Gradient (MS-KAI-CG) algorithm, whose preliminary results
have been reported in \cite{Pinto2}. In particular, MS-KAI-CG can be
considered as a knowledge-aided iterative (KAI) approach similar to
that of MS-KAI-ESPRIT using the CG algorithm. We also formulate an
MS-KAI-CG version with forward-backward spatial smoothing, denoted
as MS-KAI-CG-FB, which can deal with correlated signals. Unlike
prior KAI approaches, MS-KAI-CG and MS-KAI-CG-FB are no longer
restricted to the same number of iterations as the number of signals
$P$. We also conduct a study of the computational complexity in
terms of arithmetic operations of the proposed and previously
reported DoA estimation algorithms along with a simulation study for
scenarios with closely-spaced source signals.

This paper is structured as follows. Section II describes the signal
model, the main parameters and formulates the DoA estimation
problem. Section III introduces the proposed MS-KAI-CG algorithm,
whereas Section IV presents the proposed MS-KAI-CG-FB algorithm for
correlated source signals. Section V illustrates and discusses the
computational complexity of the proposed and existing algorithms. In
Section VI, we present and discuss the simulation results whereas
the concluding remarks are given in Section VII.
\section{Signal Model and Problem Formulation}
\label{sysmodel}

In this section we describe the signal model used to study the DoA
estimation algorithms and formulate the DoA estimation problem. Let
us assume that $\mathit{P}$ narrowband signals from far-field
sources impinge on a uniform linear array (ULA) of $\mathit{M} > \mathit{P}$ sensor elements from  directions $\bm
{\theta}=\left[ \theta_{1},\theta_{2},\ldots, \theta_P\right] ^T$. We also
consider that the sensors are equally spaced from adjacent sensors
by a distance $ d\leq\frac{\lambda_{c}}{2}$, where $\lambda_{c}$ is
the signal wavelength, and that without loss of generality, we have
\text{${\frac{-\pi}{2}\leq\theta_{1}\leq\theta_{2}\ldots
\leq\theta_P\leq \frac{\pi}{2}}$}. The $i$th data snapshot of the
$\mathit{M}$-dimensional sensor array can be organized in a vector
and modeled as
\begin{equation}
\bm x(i)=\bm A\,s(i)+\bm n(i),\qquad i=1,2,\ldots,N,
\label{model}
\end{equation}
where $\bm s(i)=[s_{1}(i),\ldots,s_{P}(i)]^T
\in\mathbb{C}^{\mathit{P\times1}}$ refers to the zero-mean source
data vector, $\bm n(i) \in\mathbb{C}^{\mathit{M \times 1}}$ is the
vector of white circular complex Gaussian noise with zero mean and
variance $\sigma_n^2$, and $N$ denotes the number of available
snapshots. The steering matrix
\begin{equation}
\bm A(\bm \Theta)=[\bm
a(\theta_{1}),\ldots,\bm a(\theta_{P})] \in\mathbb
{C}^{\mathit{M\times P}},
\label{array manifold}
\end{equation}
which is known as the array manifold and has a Vandermonde structure
contains the array steering vectors $\bm a(\theta_n)$ that
correspond to the $n$th source, which can be expressed by
\begin{equation}
\bm a(\theta_n)=[1,e^{j2\pi\frac{d}{\lambda_{c}}
    \sin\theta_n},\ldots,e^{j2\pi(M-1)\frac{d}{\lambda_{c}}\sin\theta_n}]^T,
\label{steer}
\end{equation}
where $n=1,\ldots, P$. Using the fact that $\bm s(i)$ and $\bm n(i)$
are modeled as statistically independent random variables, the
$M\times M$  covariance matrix of the received data can be
calculated as follows:
\begin{equation}
\bm R=\mathbb E\left[\bm x(i)\bm x^H(i)
\right]=\bm A\,\bm R_{ss}\bm A^H+
\sigma_n^2\bm I_M,
\label{covariance}
\end{equation}
where the superscript \textit{H} and $\mathbb E[\cdot]$ in $\bm
R_{ss}=\mathbb E[\bm s(i)\bm s^H(i)]$ and in $\mathbb E[\bm n(i)\bm
n^H(i)]=\sigma_n^2\bm I_M^{}$ denote the Hermitian transposition and
the expectation operator and $\bm I_M$ stands for the $M\times M$
identity matrix. Since the true signal covariance matrix is unknown,
it can be estimated by the sample average formula given by
\begin{equation}
\bm {\hat{R}}_{o}=\frac{1}{N} \sum\limits^{N}_{i=1}\bm x(i)\bm x^H(i),
\label{covsample}
\end{equation}
whose estimation accuracy is dependent on $N$. The problem we are
interested in solving is how to exploit prior knowledge about source
signals and the structure of ${\bm R}$ to devise high-performance
DoA estimation techniques using the proposed MS-KAI-CG and
MS-KAI-CG-FB algorithms.
\section{Proposed MS-KAI-CG Algorithm }
\label{Proposed_MS_KAI}

In this section, we present the derivation and the details of the
proposed MS-KAI-CG algorithm. Let us begin by rewriting
\eqref{covsample} using \eqref{model} as given by \cite{Vorobyov2}:
\begin{align}
\bm {\hat{R}}_{o}=&\frac{1}{N} \sum\limits^{N}_{i=1}(\bm A\,s(i)+\bm
n(i))\:(\bm A\,s(i)+\bm n(i))^H \nonumber\\=& \underbrace{\bm
A\left\lbrace\frac{1}{N} \sum\limits^{N}_{i=1}\bm s(i)\bm
s^H(i)\right\rbrace\bm A^H+\:\frac{1}{N} \sum\limits^{N}_{i=1}\bm
n(i)\bm n^H(i)}_{\text{"former part"}}\;\nonumber\\+&\underbrace{\bm A\left\lbrace\frac{1}{N}
    \sum\limits^{N}_{i=1}\bm s(i)\bm n^H(i)\right\rbrace\:
    +\:\left\lbrace\frac{1}{N} \sum\limits^{N}_{i=1}\bm n(i)\bm
    s^H(i)\right\rbrace\bm{A}^{H}}_{\text{"latter part = unwanted interference"}}
\label{expandedcovsample1}
\end{align}
The former part of the covariance matrix \text{$\bm {\hat{R}}_{o}$}
in \eqref{expandedcovsample1} can be seen as the sum of the
estimates of the two terms of \text{$\bm R$} given in
\eqref{covariance}, which account for the signal and the noise
components, respectively. The latter part in
\eqref{expandedcovsample1} can be considered  as the sum of unwanted
interference terms, which correspond to terms containing the
correlation between the signal and the noise vectors. The system
model that is studied employs noise vectors that have zero-mean and
are statistically independent of the signal vectors. Therefore, the
signal and noise components are stochastically independent of one
another. As a result, for a sufficiently large number of samples
$N$, the unwanted interference indicated in
\eqref{expandedcovsample1} tends to zero. However, in practice we
often have access to a reduced number of snapshots. In such
situations, the interference in \eqref{expandedcovsample1} that
represent unwanted quantities may become substantial and have an
impact on the estimated signal and orthogonal subspaces, which can
significantly differ from the actual signal and noise subspaces. {An
effect of this interference is the existence of part of the true
signal eigenvector inside the sample noise subspace and conversely.
This overlap between subspaces is defined as the average value of
the energy of the estimated signal eigenvectors flowed into the true
orthogonal subspace. The reduction of  this overlap as a result of
decreasing the undesirable terms applied to the specific case of
Root-MUSIC algorithm  is detailed in \cite{Vorobyov2}. Recent work
\cite{Pinto4} shows that the first of multiple possible iterations
to reduce the unwanted interference already provides a covariance
matrix whose mean squared error (MSE) is less than or equal to the
MSE of the original covariance matrix estimate. This inequality
ensures the improvement of that estimate (4) and can be viewed as
another, but sufficient reason for assessing the performances of DoA
estimation methods based on this refinement.}

The central point of the proposed MS-KAI-CG algorithm is to combine
the  depuration of the estimate of the data covariance matrix at
each iteration with the incorporation of the knowledge provided by
the more modern steering matrices which gradually include the
updated estimates from the earlier iteration. Based on these more
modern steering matrices, purer estimates of the projection matrices
of the signal and orthogonal subspaces are calculated. These
estimates of projection matrices associated with the initial
estimate of the data covariance matrix and the correction factor
employed to subtract the unwanted terms allow us to obtain improved
estimates. The best estimates are obtained by choosing among the
group of estimates the set that has the minimum value of the
correction function, i.e., the set of DoAs with the strongest
likelihood. The refined covariance matrix is computed by gradually
subtracting the unwanted terms of $\bm {\hat{R}}_{o}, $ which are
shown in \eqref{expandedcovsample1}.

The steps of the proposed MS-KAI-CG algorithm are listed in Table
\ref{Multi_Step_KAI_CG}. The MS-KAI-CG algorithm begins with the  computation of
the estimate of the data covariance matrix \eqref{covsample}. Next,
the DoAs are estimated using the CG DoA estimation algorithm
reported in \cite{Semira}.

Here, the number of signals \textit{P} or the model order is assumed
to be known, which is an assumption often adopted in the literature.
Alternatively, the number of signals \textit{P} could be estimated
by model-order selection algorithms
\cite{Rappaport,Schell,Rissanen}. The CG method, from which the
first and the last steps of the MS-KAI-CG are based on, is used to
reduce to a minimum a loss function, or analogously, to find a solution to a linear system
of equations by approaching the optimal solution gradually through a
line search along consecutive directions, which are sequentially
calculated at each direction \cite{golub}. As a result of the use of
the CG algorithm in DoA estimation, we have a system of equations
that is iteratively solved for $\bm{w}$ at each search angle as
described by
\begin{equation}
\bm{Rw}=\bm{b}(\theta),
\label{wienhopf}
\end{equation}
where $\bm{R}$ is the covariance matrix of the received data
\eqref{covariance}, which in practice must be estimated by
\eqref{covsample}, and $\bm{b}(\theta)$ is the primary vector
defined as
\begin{equation}
\bm{b}(\theta)=\frac{\bm{R}\:\bm{a}(\theta)}{\|\bm{R}\:\bm{a}(\theta )\|}
\label{inicialvec1}
\end{equation}
where $\bm{a}(\theta)$ is the search vector. The mentioned vector
has the shape of the steering vector \eqref{steer} and gradually
increases from $-90^{o}$ to  $90^{o}$.

The expanded signal subspace of rank \textit{P} is calculated by
means of the CG algorithm, which is summed up in the Table
\ref{Summary_of_CG}. For each primary vector described in
\eqref{inicialvec1}, the group of orthogonal vestigial vectors
\eqref{ortbasextsigsubs} is constructed after carrying out $
\mathrm{P} $ iterations of the CG algorithm.

\begin{equation}
\bm{G}_{cg,P+1}(\theta)=[\bm{g}_{cg,0}(\theta), \bm{g}_{cg,1}(\theta),\ldots, \bm{g}_{cg,P}(\theta)],
\label{ortbasextsigsubs}
\end{equation}
where $\bm{b}(\theta)$=$\;\bm{g}_{cg,0}(\theta)$ generates the
expanded Krylov subspace consisting of the actual signal subspace of
dimension $ \mathit{P} $ and the search vector itself. All the
vestigial vectors are normalized except for the last one. If $\theta
\in\{\theta_1,\ldots\,,\theta_P\}$, the primary vector
$\bm{b}(\theta)$ lies in the true signal subspace spanned by the
$[\bm{g}_{cg,0}(\theta), \bm{g}_{cg,1}(\theta),\ldots,
\bm{g}_{cg,P-1}(\theta)]$ basis vectors of the expanded Krylov
subspace. Thus, the rank of the generated signal subspace decreases from
$\mathit{P}+1$ to $\mathit{P} $ and we have
\begin{equation}
\bm{g}_{cg,P}(\theta)=0,
\label{rankdrop}
\end{equation}
where $\bm{g}_{cg,P}$ is the last unnormalized vestigial vector.
In order to take advantage of this property, the proposed
{MS-KAI-CG} algorithm makes use of the spectral function defined in
\cite{Grover}:
\begin{equation}
\mathcal{P_{K}}(\theta^{(n)})=\frac{1}  {\|\bm{g}^{H}_{cg,P} (\theta^{(n)}) \bm{G}_{cg,P+1} (\theta^{(n-1)})\|^2},
\label{spectralfun}
\end{equation}
where $\theta^{(n)}$ refers to the search angle in the whole angle
range $\{-90^{o},\ldots,90^{o}\}$ with $\theta^{(n)}=
n\varDelta^{o}-90^{o}$, where $\varDelta^{o}$ is the search step and
$n=0, 1,\ldots, 180^{o}/\varDelta^{o}$. The matrix $\bm{G}_{cg,P+1}
(\theta^{(n-1)})$ contains all vestigial vectors at the $(n-1)$th
%angle and $\bm{g}_{cg,P} (\theta^{(n)})$ is the last vestigial
vector calculated at the current search step $\mathit{n}$. If
$\theta^{(n)}\in\{\theta_{1},\ldots,\theta_{P}\}$, $\bm{g}_{cg,P}
(\theta^{(n)})= 0$ and we can expect a peak in the spectrum. Taking
into account that $\bm {\hat{R}}_{o}$ in \eqref{covsample} is only a
sample average estimate, which is unknown in real life,
$\bm{g}_{cg,P} (\theta^{(n)})$ and $\bm{G}_{cg,P+1}
(\theta^{(n-1)})$ become approximations. Hence the spectral function
in \eqref{spectralfun} provides large values but they do not tend to
infinity as for the original covariance matrix. It must be
highlighted that the choice of $\frac{1}  {\|\bm{g}_{cg,P}
(\theta^{(n)})\|^2} $ as a localization function, instead of
\eqref{spectralfun}, was first considered. Due to unsatisfactory
results \cite{Semira}, \eqref{spectralfun} was  preferred.
\begin{table}[!h]
    %\centering
    \small
    \caption{ Summary of the Conjugate Gradient Algorithm}\smallskip
    %\centering
    \scalebox{0.90}\medskip{
        \begin{tabular}{r l }
            \hline
            \\
            %\multicolumn{2}{|l|}{\small $\textbf{\underline{First stage}:}$}\\[0.6ex]
            \multicolumn{2}{c}{\small $\bm{w}_{0}=0,\: \bm{d}_{1}= \bm{g}_{cg,0}=b,\:\mathrm{\rho}_{0}=\bm{g}_{cg,0}^H\bm{g}_{cg,0}$}\\[0.6ex]
            \multicolumn{2}{l}{\small \textbf{for} \textit{i}=1 to \textit{P} do:} \\[0.6ex]
            {\small $\bm{v}_{i}$} & ={\small $\;\bm{R\:d}_{i}$} \\ [0.6ex]
            {\small $\mathrm{\alpha}_{i}$} & ={\small $\;\mathrm{\rho}_{i-1}\:/\:\bm{d}_{i}^H\bm{v}_{i}$} \\[0.6ex]
            {\small $\bm{w}_{i}$ }&={\small $\;\bm{w}_{i-1}$} + {\small $\bm{\alpha}_{i}\bm{d}_{i}$ } \\[0.6ex]
            {\small $\bm{g}_{cg,i}$} &={\small $\;\bm{g}_{cg,i-1}$} - {\small $\bm{\alpha}_{i}\bm{v}_{i}$} \\[0.6ex]
            {\small $\mathrm{\rho}_{i}$} & ={\small $\;\bm{g}_{cg,i}^H\bm{g}_{cg,i}$}   \\[0.6ex]

            {\small $\mathrm{\beta}_{i}$} & ={\small $\;\mathrm{\rho}_{i}\:/\:\mathrm{\rho}_{i-1}\;$}={\small $\;\|\bm{g}_{cg,i}\|^2\:/\:\|\bm{g}_{cg,i-1}\|^2$}         \\[0.6ex]

            {\small $\bm{d}_{i+1}$} &={\small $\;\bm{g}_{cg,i}\;$}+ {\small $\bm{\beta}_{i}\bm{d}_{i}$} \\[0.6ex]
            \multicolumn{2}{l}{\small \textbf{end for}} \\[0.6ex]

            \multicolumn{2}{l}{\small \textbf{form}\hspace{7mm}$\bm{G}_{cg,P+1}(\theta)$\hspace{2mm}\eqref{ortbasextsigsubs}\hspace{2mm}} \\[0.6ex]

            \multicolumn{2}{l}{\small \textbf{compute}\hspace{3mm}$\mathcal{P_{K}}(\theta^{(n)})$\hspace{2mm}\eqref{spectralfun}} \\[0.6ex]

            \multicolumn{2}{l}{\small \textbf{find}\hspace{9mm}$\hat{P}$ \text{largest peaks of} $\mathcal{P_{K}}(\theta^{(n)})$\hspace{1mm}\text{to obtain}} \\[0.6ex]

            \multicolumn{2}{l}{\hspace{15mm}\small \text{estimates}\hspace{1mm} $\hat{\theta_{\mathit{l}}}$ \text{of the DOA}}\\[0.6ex]

            \hline
        \end{tabular}
    }
    %\caption{Table to test captions and labels}
    \label{Summary_of_CG}

\end{table}

The superscript $(\cdot)^{(1)}$ refers to the estimation task
performed in the first step. Next, a procedure consisting of $n=1:I$
iterations starts by forming the array manifold with the steering
vectors \eqref{array manifold} using the DoA estimates. Then, the
amplitudes of the sources are estimated such that the square norm of
the differences between the vector of observations and the vector
containing estimates and the available known DoAs is minimized. This
problem can be expressed \cite{Vorobyov2} as:
\begin{eqnarray}
\hat{\bm{s}}(i)=\arg\min_{\substack{\bm
        s}}\parallel\bm{x}(i)-\hat{\bm{A}}\bm{s}\parallel^2_2.
\label{minimization01}
\end{eqnarray}
The minimization of \eqref{minimization01} is achieved using the widely known
LS method \cite{Rappaport} and the solution is described by
\begin{equation}
\hat{\bm{s}}(i)=(\bm{\hat{A}}^{H}\:\bm{\hat{A}})^{-1}\:\bm{\hat{A}}^{H}\:\bm{x}(i)
\label{minimization2}
\end{equation}
The noise component is then computed as the difference between the
estimated signal and the observations made by the array, as given by
\begin{eqnarray}
\hat{\bm n}(i)=\bm x(i)\:-\: \hat{\bm A}\:\hat{\bm s}(i).
\label{noise_component}
\end{eqnarray}
After estimating the signal and noise vectors, the third term in
\eqref{expandedcovsample1} can be computed as
\begin{align}
\bm{V}&\triangleq \hat{\bm{A}}\left\lbrace\frac{1}{N}
\sum\limits^{N}_{i=1}\bm \hat{\bm{s}}(i)\bm
\hat{\bm{n}}^H(i)\right\rbrace\nonumber\\&=\hat{\bm{A}}\left\lbrace\frac{1}{N}
\sum\limits^{N}_{i=1}(\bm{\hat{A}}^{H}\:\bm{\hat{A}})^{-1}\bm{\hat{A}}^{H}\bm{x}(i)\right.\nonumber\\&\left.\times(\bm{x}^{H}(i)-\bm{x}^{H}(i)\hat{\bm{A}}(\hat{\bm{A}}^{H}\hat{\bm{A}})^{-1}\:\hat{\bm{A}}^{H})\right\rbrace\nonumber\\&=\bm{\hat{Q}}_{A}\left\lbrace\frac{1}{N}
\sum\limits^{N}_{i=1}
\bm{x}(i)\bm{x}^H(i)\:\left(\bm{I}_{M}\:-\:\hat{\bm{Q}}_{A}\right)
\right\rbrace\nonumber\\&=\bm{\hat{Q}}_{A}\:\bm{\hat{R}}\:\bm{\hat{Q}}_{A}^{\perp},
\label{terms_deducted}
\end{align}
where
\begin{equation}
\bm{\hat{Q}}_{A}\triangleq
\bm{\hat{A}}\:(\bm{\hat{A}}^{H}\:\bm{\hat{A}})^{-1}\:\bm{\hat{A}}^{H}
\end{equation}
is an estimate of the projection matrix of the signal subspace, and
\begin{equation}
\bm{\hat{Q}}_{A}^{\perp}\triangleq\bm{I}_{M}\:-\:\bm{\hat{Q}}_{A}
\end{equation}
is an estimate of the projection matrix of the orthogonal subspace.

Next, as part of the procedure with $n=1:I $ iterations, the depurated
data covariance matrix $\bm{\hat{R}}^{(n+1)}$ is obtained by
computing an improved version of the estimated terms from the
initial data covariance matrix as given
\begin{equation}
\label{modified_data_covariance} \bm{\hat{R}}^{(n+1)} =
\bm{\hat{R}}_{o}\:-\:\mathrm{\mu}\:(\bm{V}^{(n)}\:+\:\bm{V}^{(n)H}),
\end{equation}
where the superscript $(\cdot)^{(n)}$ refers to the $n^{th} $
iteration performed. The correction factor \text{$\mu $} increases
from 0 to 1 incrementally, resulting in refined data covariance
matrices. Each of them is the basis for estimating  new DoAs also
denoted by the superscript  $(\cdot)^{(n+1)}$ by using the CG
algorithm, which was previously described. Then, a new matrix with
the steering vectors $\bm{\hat{B}}^{(n+1)}$ is formed by the
steering vectors of those newer estimated DoAs. By using this new
matrix, it is possible to calculate the newer estimates of the
projection matrices of the signal \text{$
    \bm{\hat{Q}}_{B}^{(n+1)} $} and the orthogonal \text{$
    \bm{\hat{Q}}_{B}^{(n+1)\perp} $} subspaces.
Subsequently, employing the newer estimates of the projection
matrices, the initial sample data matrix, $\bm {\hat{R}}_{o}$
\eqref{covsample}, and the number of sensors and sources, the
statistical correction function $\mathit{U^{(n+1)}(\mu)}$ \cite{Stoica} is
calculated for each value of \text{$\mu $} at the $n^{th}$ iteration,
as follows:
\begin{equation}
\mathit{U^{(n+1)}(\mu)}=\mathrm{ln\:det}\left(\cdot\right),
\label{SML_objective_function_1}
\end{equation}
where
\begin{equation}
\left(\cdot\right)
=\left(\bm{\hat{Q}}_{B}^{(n+1)}\:\bm{\hat{R}}_{o}\:\bm{\hat{Q}}_{B}^{(n+1)}+\dfrac{{\rm
        Trace}\{\bm{\hat{Q}}_{B}^{\perp\:(n+1)}\:\bm{\hat{R}}_{o}\}}
{\mathit{M-P}}\:\bm{\hat{Q}}_{B}^{\:(n+1)\perp}\right)\nonumber
\end{equation}
The earlier calculation allows choosing the group of unavailable DoA
estimates that have a stronger likelihood at each iteration. Then,
the group of estimated DoAs corresponding to the optimal value of
\text{$\mu $} that reduce \eqref{SML_objective_function_1} to a minimum also
at each $n^{th}$ iteration is determined. Lastly, the output of the
MS-KAI-CG algorithm corresponds to the group of DoA estimates
determined at the $I^{th}$ iteration, as described in Table
\ref{Multi_Step_KAI_CG}.

\begin{table}[htb]
    %\centering
    \small
    \caption{Proposed MS-KAI-CG Algorithm}\smallskip
    %\vspace{2mm}
    %\centering
    \scalebox{0.90}{
        \begin{tabular}{r l}
            \hline\\

            \multicolumn{2}{l}{\small $\textbf{\underline{Inputs}:}$}\\[0.7ex]
            \multicolumn{2}{l}{\small$\mathit{M}$,\hspace{2mm}$\mathit{d}$,\hspace{2mm}$\lambda$,\hspace{2mm}$\mathit{N}$,\hspace{2mm}$\mathit{I}$ }\\[0.6ex]
            \multicolumn{2}{l}{\small\text{Received vectors}  $\bm x(1)$,\hspace{2mm}$\bm x(2)$,$\cdots$, $\bm x(N)$}\\[0.6ex]

            \multicolumn{2}{l}{\small $\textbf{\underline{Outputs}:}$}\\[0.6ex]

            \multicolumn{2}{l}{\small\text{Estimates}\hspace{1mm}$\mathit{\hat{\theta}_{1}^{(n+1)}(\mu\,opt)}$,\hspace{2mm}$\mathit{\hat{\theta}_{2}^{(n+1)}(\mu\,opt)}$,$\cdots$,\hspace{2mm}$\mathit{\hat{\theta}_{P}^{(n+1)}(\mu\,opt)}$} \\[3.1ex]
            %\hline

            \multicolumn{2}{l}{\small $\textbf{\underline{First step}:}$}\\[0.9ex]

            \multicolumn{2}{l}{\small $\bm{\hat{R}}_{o}=\frac{1}{N} \sum\limits^{N}_{i=1}\bm x(i)\bm x^H(i)$}\\[1.8ex]

            \multicolumn{2}{l}{\small $\{\mathit{\hat{\theta}_{1}}^{(1)},\:\mathit{\hat{\theta}_{2}}^{(1)},\cdots,\mathit{\hat{\theta}_{P}}^{(1)}\}\;\;\underleftarrow{CG}$ $\:(\bm{\hat{R}}_{o},P,d,\lambda)$}\\[1.2ex]

            \multicolumn{2}{l}{\small$\bm{\hat{A}}^{(1)}=\left[\bm{a}(\mathit{\hat{\theta}_{1}^{(1)}}),\bm{a}(\mathit{\hat{\theta}_{2}^{(1)}}),\cdots,\bm{a}(\mathit{\hat{\theta}_{P}^{(1)}})\right]$} \\ [1.1ex]

            \multicolumn{2}{l}{\small $\textbf{\underline{Second step}:}$}\\[0.9ex]

            \multicolumn{2}{l}{\small \textbf{for}  \textit{n}\hspace{1mm}=\hspace{2mm}\text{1}\hspace{1mm}:\hspace{1mm}\textit{I}}\\[0.6ex]

            \multicolumn{2}{l}{\small $\bm{\hat{Q}}_{A}^{(n)}= \bm{\hat{A}}^{(n)}\:(\bm{\hat{A}}^{(n)H}\:\bm{\hat{A}}^{(n)})^{-1}\:\bm{\hat{A}}^{(n)H}$ }\\[0.9ex]

            \multicolumn{2}{l}{\small$\bm{\hat{Q}}_{A}^{(n)\perp}=\bm{I}_{M}\:-\:\bm{\hat{Q}}_{A}^{(n)}$ }\\[1.0ex]

            \multicolumn{2}{l}{\small $\bm{V}^{(n)}=\bm{\hat{Q}}_{A}^{(n)}\:\bm{\hat{R}}_{o}\:\bm{\hat{Q}}_{A}^{(n)\perp}$}\\[1.8ex]

            %            \multicolumn{2}{l}{\small \textbf{for} $\bm{\mu=}\hspace{1mm}\text{0}:\text{increment\hspace{1mm}:\hspace{1mm}1} $}\\[0.9ex]

             \multicolumn{2}{l}{\small \textbf{for} $\bm{\mu}=\mathrm{\:0:\textcolor{red}{\iota}\::1}$}\\[0.9ex]

            \multicolumn{2}{l}{\small $ \bm{\hat{R}}^{(n+1)} = \bm{\hat{R}}_{o}\:-\:\mathrm{\mu}\:(\bm{V}^{(n)}\:+\:\bm{V}^{(n)H})$} \\[0.9ex]

            \multicolumn{2}{l}{\small $\{\mathit{\hat{\theta}_{1}}^{(n+1)},\:\mathit{\hat{\theta}_{2}}^{(n+1)},\cdots,\mathit{\hat{\theta}_{P}}^{(n+1)}\}\;\;\underleftarrow{CG}$ $\:(\bm{\hat{R}}^{(n+1)},\:P,d,\lambda)$}\\[1.4ex]

            \multicolumn{2}{l}{\small$\bm{\hat{B}}^{(n+1)}=\left[\bm{a}(\mathit{\hat{\theta}_{1}^{(n+1)}}),\bm{a}(\mathit{\hat{\theta}_{2}^{(n+1)}}),\cdots,\bm{a}(\mathit{\hat{\theta}_{P}^{(n+1)}})\right]$} \\ [1.4ex]

            \multicolumn{2}{l}{\small $\bm{\hat{Q}}_{B}^{(n+1)}= \bm{\hat{B}}^{(n+1)}\:(\bm{\hat{B}}^{(n+1)H}\:\bm{\hat{B}}^{(n+1)})^{-1}\:\bm{\hat{B}}^{(n+1)H}$}\\[1.2ex]

            \multicolumn{2}{l}{\small$\bm{\hat{Q}}_{B}^{(n+1)\perp}=\bm{I}_{M}\:-\:\bm{\hat{Q}}_{B}^{(n+1)}$}\\[1.2ex]

            \multicolumn{2}{l}{\small $\mathit{U^{(n+1)}(\mu)}=\mathrm{ln\:det}\left(\cdot\right),$}\\[1.1ex]

            \multicolumn{2}{l}{\small$\left(\cdot\right) =\left(\bm{\hat{Q}}_{B}^{(n+1)}\:\bm{\hat{R}}_{o}\:\bm{\hat{Q}}_{B}^{(n+1)}+\dfrac{{\rm Trace}\{\bm{\hat{Q}}_{B}^{\perp\:(n+1)}\:\bm{\hat{R}}_{o}\}} {\mathit{M-P}}\:\bm{\hat{Q}}_{B}^{\:(n+1)\perp}\right)$}\\[1.1ex]

            \multicolumn{2}{l}{\small $\mathit{\mu}_{\mathrm{o}}^{(n+1)}=\arg \min \hspace{1mm}\mathit{U^{(n+1)}(\mu)}$}\\[1.0ex]

            \multicolumn{2}{l}{\small $\mathrm{DoAs}^{(n+1)}= \left(\color{red}\$\right),$}\\[1.1ex]

            \multicolumn{2}{l}{\small$\left(\color{red}\$\right) =
                \{\mathit{\hat{\theta}_{1}^{(n+1)}(\mu_{o})}$,\hspace{2mm}$\mathit{\hat{\theta}_{2}^{(n+1)}(\mu_{o})}$,$\cdots$,\hspace{2mm}$\mathit{\hat{\theta}_{P}^{(n+1)}(\mu_{o})}\}$}\\[1.0ex]

            \multicolumn{2}{l}{\small$ \textbf{if}\hspace{1mm}  \mathit{n<=\;P}$}\\[0.6ex]

            \multicolumn{2}{l}{\small $\bm{\hat{A}}^{(n+1)}=\left\{\bm{a}(\mathit{\hat{\theta}_{\{1,\cdots,n\}}^{(n+1)}(\mu_{o})})\right\}\bigcup\left\{\bm{a}(\mathit{\hat{\theta}_{\{1,\cdots,P\}\,-\,\{1,\cdots,n\}}^{(1)}})\right\}$}\\[1.0ex]

            \multicolumn{2}{l}{\small \textbf{else}}\\[1.0ex]

            \multicolumn{2}{l}{\small$\bm{\hat{A}}^{(n+1)}= \left(\color{red}\#\right),$}\\[1.1ex]

            \multicolumn{2}{l}{\small$\left(\color{red}\#\right) =
                \left[\bm{a}(\mathit{\hat{\theta}_{1}^{(n+1)}}(\mu_{o})),\bm{a}(\mathit{\hat{\theta}_{2}^{(n+1)}}(\mu_{o})),\cdots,
                \bm{a}(\mathit{\hat{\theta}_{P}^{(n+1)}}(\mu_{o}))\right]$} \\ [1.4ex]

            \multicolumn{2}{l}{\small \textbf{end if}}\\[1.0ex]
            \multicolumn{2}{l}{\small \textbf{end for}}\\[1.0ex]

            \multicolumn{2}{l}{\small \textbf{end for}}\\[1.0ex]

            \hline

        \end{tabular}
    }
    %\caption{Table to test captions and labels}
    \label{Multi_Step_KAI_CG}

\end{table}

\section{Proposed {MS-KAI-CG-FB} Algorithm }

DoA estimation algorithms often experience performance degradation
in the presence of correlated signals. This issue has been verified
for the proposed MS-KAI-CG algorithm, as will be shown later on. In
this section, we present an approach that combines the proposed
MS-KAI-CG algorithm and the well-known forward-backward spatial
smoothing (FBSS) \cite{Pillai,Haardt} technique, referred to as
MS-KAI-CG-FB algorithm, for dealing with correlated signals. In the
proposed MS-KAI-CG-FB algorithm, the FBSS covariance matrix
\eqref{FB_smoothed} is obtained from the initial estimate of the
data covariance matrix $\bm {\hat{R}}_{o}$ \eqref{covsample}, as
follows:
\begin{equation}
\bm {\hat{R}}=\frac{1}{K} \sum\limits^{K}_{k=1}\bm
{Z}_{k}\:\bm{\tilde{R}}\:\bm{Z} _{k}^{T},
\label{FB_smoothed}
\end{equation}
where the number of subarrays employed is obtained by
\begin{equation}
\mathit{K=M-L+1}
\label{Number_of_subarrays_FB},
\end{equation}
In \eqref{Number_of_subarrays_FB}, the parameter $\textit{L}$ refers
to the number of sensors of the subarrays and $\textit{M}$
designates the number of sensors of the original ULA. The matrix
$\bm{Z} _{k}$ is given by
\begin{equation}
\bm{Z} _{k}=\left[\bm{0}_{L \times(k-1)}\;|\:\bm{I}_{L \times L}\;|\:\bm{0}_{L \times M-(L+k-1)}\;\right]
\label{Matrix_of_subarrays_FB}
\end{equation}
The forward-backward refined matrix $\bm{\tilde{R}}$ is defined as
\begin{equation}
\bm {\tilde{R}}=\frac{1}{2}\left(\bm {\hat{R}}_{o}\:+\: \bm {J}\:\bm {\hat{R}}_{o}^{*}\:\bm {J}\right),
\label{Matrix_FB}
\end{equation}
where $\bm {J}$ is an reversal matrix described by
\begin{equation}
\bm {J} =
\begin{bmatrix}
0 && 1\\
&\reflectbox{$\ddots$} \\
1& & 0
\end{bmatrix},
\end{equation}
and $\left( *\right)$ denotes complex conjugate.

Next, we rewrite \eqref{FB_smoothed} using \eqref{model} as follows:
\begin{align}
\bm {\hat{R}}=&\frac{1}{N} \sum\limits^{N}_{i=1}(\bm A\,s(i)+\bm
n(i))\:(\bm A\,s(i)+\bm n(i))^H \nonumber\\=& \underbrace{\bm
A\left\lbrace\frac{1}{N} \sum\limits^{N}_{i=1}\bm s(i)\bm
s^H(i)\right\rbrace\bm A^H+\:\frac{1}{N} \sum\limits^{N}_{i=1}\bm
n(i)\bm n^H(i)}_{\text{"former part"}}\nonumber\\+&\underbrace{\bm A\left\lbrace\frac{1}{N}
    \sum\limits^{N}_{i=1}\bm s(i)\bm n^H(i)\right\rbrace\:
    +\:\left\lbrace\frac{1}{N} \sum\limits^{N}_{i=1}\bm n(i)\bm
    s^H(i)\right\rbrace\bm{A}^{H}}_{\text{"latter part = unwanted interference"}}
\label{expandedcovsample}
\end{align}
Similarly to \eqref{expandedcovsample1}, in section
\ref{Proposed_MS_KAI}, the former part of \text{$\bm {\hat{R}}$} in
\eqref{expandedcovsample} can be viewed as the sum of the estimates
of the two terms of \text{$\bm R$} given in \eqref{covariance},
which represent the signal and the noise components, respectively.
Following the same reasoning, the latter part of
\eqref{expandedcovsample},  can  be seen as unwanted interference,
i.e., cross correlated terms that tend to zero for a large enough
number of samples.
%The system model under
%study is based on noise vectors which are zero-mean  and also
%independent of the signal vectors. Thus, the signal and noise
%components are uncorrelated to each other. As a consequence, for a
%large enough number of samples $N$, the last two  terms pointed out in
%\eqref{expandedcovsample} tend to zero. Nevertheless, in practice
%the number of available samples can be limited. In such situations,
However, the unwanted interference in \eqref{expandedcovsample} may have
large values, which result in estimates of the signal and the
orthogonal subspaces that deviate from the actual subspaces.

The key advantage of the suggested MS-KAI-CG-FB algorithm is to
refine the FBSS covariance matrix estimate $\bm {\hat{R}}$
\eqref{FB_smoothed} at each iteration by gradually including the
knowledge provided by the newer steering matrices which
progressively incorporate updated estimates from the prior
iteration. Based on these more modern steering matrices, refined
estimates of the projection matrices of the signal and the noise
subspaces are calculated. These estimates of projection matrices
associated with the estimate of the FBSS data covariance matrix $\bm
{\hat{R}}$ and the correction factor employed to reduce its unwanted
interference allows determining the group of estimates that has the
minimum value of the correction function, i.e., the topmost
likelihood of being the group of the actual DoAs. The refined data
covariance matrix is calculated by gradually reducing the unwanted
terms of $\bm {\hat{R}}$, which are indicated in
\eqref{expandedcovsample}.

The steps of the suggested MS-KAI-CG-FB algorithm are listed in
Table \ref{Multi_Step_KAI_CG_FB}. The algorithm begins with the calculation of
the initial sample data covariance matrix
\eqref{covsample}. Then, the FBSS covariance matrix estimate
\eqref{FB_smoothed} is determined. Subsequently, the DoAs are
estimated using the ordinary CG algorithm which was briefly
described in Section \ref{Proposed_MS_KAI}. The superscript $(\cdot)^{(1)}$
refers to the estimation task performed in the first step. Next, a
procedure consisting of $n=1:I$ iterations starts by forming the
array manifold with the steering vectors \eqref{array manifold}
using the DoA estimates. Then, the amplitudes of the sources are
estimated such that the square norm of the differences between the
vector of observations and the vector containing estimates and the
available known DoAs is minimized. This problem can be formulated in
a similar way to that described from \eqref{minimization01} to the
end of Section \ref{Proposed_MS_KAI}.

\begin{table}[!h]
    %\centering
    \small
    \caption{Proposed MS-KAI-CG-FB Algorithm}\smallskip
    %\vspace{2mm}
    %\centering
    \scalebox{0.90}{
        \begin{tabular}{r l}
            \hline\\
            \multicolumn{2}{l}{\small $\textbf{\underline{Inputs}:}$}\\[0.7ex]
            \multicolumn{2}{l}{\small$\mathit{M}$,\hspace{2mm}$\mathit{d}$,\hspace{2mm}$\lambda$,\hspace{2mm}$\mathit{N}$,\hspace{2mm}$\mathit{P}$ }\\[0.6ex]
            \multicolumn{2}{l}{\small\text{Received vectors}  $\bm x(1)$,\hspace{2mm}$\bm x(2)$,$\cdots$, $\bm x(N)$}\\[0.6ex]

            \multicolumn{2}{l}{\small $\textbf{\underline{Outputs}:}$}\\[0.6ex]

            \multicolumn{2}{l}{\small\text{Estimates}\hspace{1mm}$\mathit{\hat{\theta}_{1}^{(n+1)}(\mu\,opt)}$,\hspace{2mm}$\mathit{\hat{\theta}_{2}^{(n+1)}(\mu\,opt)}$,$\cdots$,\hspace{2mm}$\mathit{\hat{\theta}_{P}^{(n+1)}(\mu\,opt)}$} \\[2.0ex]
            %\hline

            \multicolumn{2}{l}{\small $\textbf{\underline{First step}:}$}\\[0.9ex]

            \multicolumn{2}{l}{\small $\bm{\hat{R}_{o}}=\frac{1}{N} \sum\limits^{N}_{i=1}\bm x(i)\bm x^H(i)$}\\[1.8ex]

            \multicolumn{2}{l}{\small $\bm{\hat{R}}\;\;\underleftarrow{FBSS}$ $\;\bm{\hat{R}_{o}}$}\\[1.2ex]

            \multicolumn{2}{l}{\small $\{\mathit{\hat{\theta}_{1}}^{(1)},\:\mathit{\hat{\theta}_{2}}^{(1)},\cdots,\mathit{\hat{\theta}_{P}}^{(1)}\}\;\;\underleftarrow{CG}$ $\:(\bm{\hat{R}},P,d,\lambda)$}\\[1.2ex]

            \multicolumn{2}{l}{\small$\bm{\hat{A}}^{(1)}=\left[\bm{a}(\mathit{\hat{\theta}_{1}^{(1)}}),\bm{a}(\mathit{\hat{\theta}_{2}^{(1)}}),\cdots,\bm{a}(\mathit{\hat{\theta}_{P}^{(1)}})\right]$} \\ [1.1ex]

            \multicolumn{2}{l}{\small $\textbf{\underline{Second step}:}$}\\[0.9ex]

            \multicolumn{2}{l}{\small \textbf{for}  \textit{n}\hspace{1mm}=\hspace{2mm}\text{1}\hspace{1mm}:\hspace{1mm}\textit{I}}\\[0.6ex]

            \multicolumn{2}{l}{\small $\bm{\hat{Q}}_{A}^{(n)}= \bm{\hat{A}}^{(n)}\:(\bm{\hat{A}}^{(n)H}\:\bm{\hat{A}}^{(n)})^{-1}\:\bm{\hat{A}}^{(n)H}$ }\\[0.9ex]

            \multicolumn{2}{l}{\small$\bm{\hat{Q}}_{A}^{(n)\perp}=\bm{I}_{M}\:-\:\bm{\hat{Q}}_{A}^{(n)}$ }\\[1.0ex]

            \multicolumn{2}{l}{\small $\bm{V}^{(n)}=\bm{\hat{Q}}_{A}^{(n)}\:\bm{\hat{R}}\:\bm{\hat{Q}}_{A}^{(n)\perp}$}\\[1.8ex]

            \multicolumn{2}{l}{\small \textbf{for} $\bm{\mu}=\mathrm{\:0:\iota\::1}$}\\[0.9ex]

            \multicolumn{2}{l}{\small $ \bm{\hat{R}}^{(n+1)} = \bm{\hat{R}}\:-\:\mathrm{\mu}\:(\bm{V}^{(n)}\:+\:\bm{V}^{(n)H})$} \\[0.9ex]

            \multicolumn{2}{l}{\small $\{\mathit{\hat{\theta}_{1}}^{(n+1)},\:\mathit{\hat{\theta}_{2}}^{(n+1)},\cdots,\mathit{\hat{\theta}_{P}}^{(n+1)}\}\;\;\underleftarrow{CG}$ $\:(\bm{\hat{R}}^{(n+1)},\:P,d,\lambda)$}\\[1.4ex]

            \multicolumn{2}{l}{\small$\bm{\hat{B}}^{(n+1)}=\left[\bm{a}(\mathit{\hat{\theta}_{1}^{(n+1)}}),\bm{a}(\mathit{\hat{\theta}_{2}^{(n+1)}}),\cdots,\bm{a}(\mathit{\hat{\theta}_{P}^{(n+1)}})\right]$} \\ [1.4ex]

            \multicolumn{2}{l}{\small $\bm{\hat{Q}}_{B}^{(n+1)}= \bm{\hat{B}}^{(n+1)}\:(\bm{\hat{B}}^{(n+1)H}\:\bm{\hat{B}}^{(n+1)})^{-1}\:\bm{\hat{B}}^{(n+1)H}$}\\[1.2ex]

            \multicolumn{2}{l}{\small$\bm{\hat{Q}}_{B}^{(n+1)\perp}=\bm{I}_{M}\:-\:\bm{\hat{Q}}_{B}^{(n+1)}$}\\[1.2ex]

            \multicolumn{2}{l}{\small $\mathit{U^{(n+1)}(\mu)}=\mathrm{ln\:det}\left(\cdot\right),$}\\[1.1ex]

            \multicolumn{2}{l}{\small$\left(\cdot\right) =\left(\bm{\hat{Q}}_{B}^{(n+1)}\:\bm{\hat{R}}\:\bm{\hat{Q}}_{B}^{(n+1)}+\dfrac{{\rm Trace}\{\bm{\hat{Q}}_{B}^{\perp\:(n+1)}\:\bm{\hat{R}}\}} {\mathit{L-P}}\:\bm{\hat{Q}}_{B}^{\:(n+1)\perp}\right)$}\\[1.1ex]

            \multicolumn{2}{l}{\small $\mathit{\mu}_{\mathrm{o}}^{(n+1)}=\arg \min \hspace{1mm}\mathit{U^{(n+1)}(\mu)}$}\\[1.0ex]

            \multicolumn{2}{l}{\small $\mathrm{DoAs}^{(n+1)}= \left(\color{red}\$\right),$}\\[1.1ex]

            \multicolumn{2}{l}{\small$\left(\color{red}\$\right) =
                \{\mathit{\hat{\theta}_{1}^{(n+1)}(\mu_{o})}$,\hspace{2mm}$\mathit{\hat{\theta}_{2}^{(n+1)}(\mu_{o})}$,$\cdots$,\hspace{2mm}$\mathit{\hat{\theta}_{P}^{(n+1)}(\mu_{o})}\}$}\\[1.0ex]

            \multicolumn{2}{l}{\small$ \textbf{if}\hspace{1mm}  \mathit{n<=\;P}$}\\[0.6ex]

            \multicolumn{2}{l}{\small $\bm{\hat{A}}^{(n+1)}=\left\{\bm{a}(\mathit{\hat{\theta}_{\{1,\cdots,n\}}^{(n+1)}(\mu_{o})})\right\}\bigcup\left\{\bm{a}(\mathit{\hat{\theta}_{\{1,\cdots,P\}\,-\,\{1,\cdots,n\}}^{(1)}})\right\}$}\\[1.0ex]

            \multicolumn{2}{l}{\small \textbf{else}}\\[1.0ex]

            \multicolumn{2}{l}{\small$\bm{\hat{A}}^{(n+1)}= \left(\color{red}\#\right),$}\\[1.1ex]

            \multicolumn{2}{l}{\small$\left(\color{red}\#\right) =
                \left[\bm{a}(\mathit{\hat{\theta}_{1}^{(n+1)}}(\mu_{o})),\bm{a}(\mathit{\hat{\theta}_{2}^{(n+1)}}(\mu_{o})),\cdots,
                \bm{a}(\mathit{\hat{\theta}_{P}^{(n+1)}}(\mu_{o}))\right]$} \\ [1.4ex]

            \multicolumn{2}{l}{\small \textbf{end if}}\\[1.0ex]
            \multicolumn{2}{l}{\small \textbf{end for}}\\[1.0ex]

            \multicolumn{2}{l}{\small \textbf{end for}}\\[1.0ex]

            \hline
        \end{tabular}
    }
    %\caption{Table to test captions and labels}
    \label{Multi_Step_KAI_CG_FB}

\end{table}

\section{Computational Complexity Analysis}
\label{computational_analysis}

In this section, we evaluate the computational cost of the suggested
MS-KAI-CG \cite{Pinto3} and MS-KAI-CG-FB algorithms which are
compared to the following classical subspace methods: ESPRIT
\cite{Roy}, MUSIC \cite{schimdt}, Root-MUSIC \cite{Barabell},
Conjugate Gradient (CG) \cite{Semira}, Auxiliary Vector Filtering
(AVF) \cite{Grover} and TS-ESPRIT \cite{Pinto}. The ESPRIT and
MUSIC-based methods use the Eigen Value Decomposition (EVD) of the
sample covariance matrix \eqref{covsample}. The computational burden
of MS-KAI-CG/MS-KAI-CG-FB in terms of number of multiplications is
depicted in Table \ref{Comput_Complexity1}, where
$\mathrm{\tau}=\frac{1}{ \iota} +1$ {is the number of equally spaced
points needed to limit the increments which form the range
$\mathrm{\mu}\in\left[0\;\: 1\right]  $ . The increment
$\mathrm{\iota}\in\left( 0\;\: 1\right] $, as defined in Tables
\ref{Multi_Step_KAI_CG} and \ref{Multi_Step_KAI_CG_FB}. Despite
$\mathrm{\tau}\in\left[2\;\: \infty\right) $, there are  two
following points to be considered: First, small values of
$\mathrm{\tau}$ lead to innacurate estimates, since its resulting
grid is not dense enough do provide precise values of the optimal
correction factor, which is responsible for the computation of the
DoAs at the  $n^{th}$ iteration. Second, large values of
$\mathrm{\tau}$ yield heavy computational burdens. Typically checked
values of $\mathrm{\tau}$, which  are neither too small to result in
inaccurate estimates nor too large to produce excessive
computational burdens and yield good results, lie in the range from
$\mathrm{\tau}=11$ to $\mathrm{\tau}=17$, corresponding to the range
from  $\mathrm{\iota}=0.1$ to $\mathrm{\iota}=0.0625$,
respectively.}

Considering the number of multiplications, it can be seen that for
the specific configuration used in one of the simulations described
in Section \ref{simulations} - i.e., $ P=2 $, $ M=12 $, $ N=100, $
MS-KAI-CG and MS-KAI-CG-FB show a relatively high computational
burden with
\begin{align}
\mathcal{O}\mathit{(P\tau\left[\frac{180}{\varDelta}\left(M^2\left(
    P+1\right)+M\left(6P+2 \right)   \right)  \right] )}.
\end{align}
 Similarly, the number of additions reaches
\begin{align}
\mathcal{O}\mathit{(P\tau\left[\frac{180}{\varDelta}\left(M^2\left(
    P+1\right)+M\left(5P+1 \right) \right)  \right] )}.
\end{align}
Despite the dense discretized grid yielded by a search step  equal
to $\mathit{0.2^{o}}$ which is employed in MS-KAI-CG and
MS-KAI-CG-FB algorithms, it is possible to reduce the inherent heavy
computational burdens of these algorithms by using approaches such
as that described in \cite{Steinwandt3}. The mentioned   technique
makes use of the a priori knowledge of a sub-band of the entire
angle range and then  focus the search in that specific region.

By examining the expressions for multiplications and additions for
the proposed MS-KAI-CG and MS-KAI-CG-FB algorithms, it can be seen
that the number of multiplications required by MS-KAI-CG and
MS-KAI-CG-FB is higher than the number of additions and serves as an
appropriate indicator of the computational
cost of the proposed and existing algorithms. % in Table
%\ref{Comput_Complexity1}, it can also be seen that the number of
%multiplications required by the su algorithms is more
%significant than the number of additions.
For this reason, in Table \ref{Comput_Complexity1}, we consider the
computational burden of the algorithms in terms of multiplications
for the purpose of comparisons. In that table, $\varDelta$  stands
for the search step.

\begin{figure}[!h]

    \centering % para centralizarmos a figura
    \includegraphics[width=8cm,height=6cm]{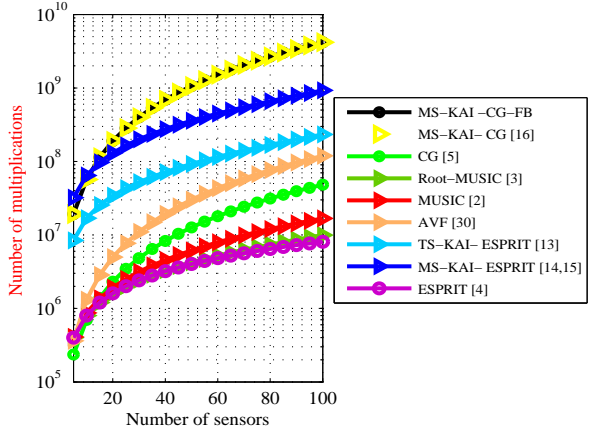} % leia abaixo
    %\vspace{-1.0em}
    \caption{\textit{Number of multiplications as powers of 10 versus number of sensors  for} $P=4$, $N=100$.}

    \label{figura:Multiplications_powers_of_10}
\end{figure}

Next, based on Table \ref{Comput_Complexity1}, we have evaluated the
influence of the number of sensor elements on the number of
multiplications based on the specific configuration composed of
$P=4$ narrowband signals impinging  on a ULA of $M$ sensor elements
and $N=100$ available snapshots. In Fig.
\ref{figura:Multiplications_powers_of_10}, we can see the main
trends in terms of computational cost measured in multiplications of
the suggested and analyzed algorithms. By examining Fig.
\ref{figura:Multiplications_powers_of_10}, it can be noticed that in
the whole range $M=\left[ 0\ 100\right] $ of sensors, the curves
describing the exact number of multiplications computed in
MS-KAI-CG-FB and MS-KAI-CG has been merged, which translates into
equivalent burdens in terms of this kind of operation. It can also
be noticed that in the range $M=\left[ 5\ 20\right] $, MS-KAI-CG-FB,
MS-KAI-CG and MS-KAI-ESPRIT require a similar cost.

\begin{table}[!h]
    \caption{Computational complexity in terms of numbers of multiplications }
    %\vspace{2mm}
    %\centering
    \scalebox{0.92}{
        \begin{tabular}{l l p{6cm}|}       \hline
            %& \underline{Multiplications} \\
            \rule{0pt}{3ex}  & $P\:\mathit{\tau \{\frac{180}{\varDelta}[M^{2}(P+1)+M(6P+2)+P+1]}$\\[1pt]
            & $\mathit{+\frac{10}{3}M^{3}+M^{2}(N +P+3)+M(\frac{3}{2}P^{2}+\frac{1}{2}P)}$  \\[1pt]
            MS-KAI-CG &$\mathit{+P^{2}(\frac{1}{2}P+\frac{3}{2})\}}$ \\[1pt]
            $\;\; \sim $ &$\mathit{+P\:[2M^{3}+M^{2}(P)+M(\frac{1}{2}P)+P^{2}(\frac{P}{2}+\frac{3}{2}) ]}$\\[1pt]
            MS-KAI-CG&
            $\mathit{+\frac{180}{\varDelta}[M^{2}(P+1)+M(6P+2)+P+1]}$\\[1pt]
            -FB& $\mathit{+M^{2}(N+2)+MP}$\\[2pt]

            % & \underline{Additions} \\[2pt]
            %
            %        \rule{0pt}{3ex}  & $P\:\mathit{\tau \{\frac{180}{\Delta}[M^{2}(P+1)+M(5P+1)-3P-2]}$\\[1pt]
            %        & $\mathit{+\frac{10}{3}M^{3}+M^{2}(N +P-1)+M(\frac{3}{2}P^{2}+\frac{5}{2}P+1)}\}$  \\[1pt]
            %        &$\mathit{+P\:[2M^{3}-M^{2}(P-2)+M(\frac{3}{2}P^2-\frac{P}{2})-P^2-\frac{P}{2} ]}$\\[1pt]
            %        &       $\mathit{+\frac{180}{\Delta}[M^{2}(P+1)+M(5P+1)-3P-2]+M^{2} N }$\\[2pt]
            \\
            MUSIC \cite{schimdt} & $\mathit{\frac{180}{\varDelta}[M^{2}+M(2-P)-P]+8MN^{2}}$  \\[4pt]
            %\hline
            Root-MUSIC \cite{Barabell} & $\mathit{2M^{3}-M^{2}P+8MN^{2}}$  \\[4pt]
            %\hline
            AVF \cite{Grover} & $\mathit{\frac{180}{\varDelta}[M^{2}(3P+1)+M(4P-2)+P+2]}$\\[1pt]
            &$\mathit{+M^{2}N}$  \\[4pt]
            %\hline
            CG \cite{Semira} & $\mathit{\frac{180}{\varDelta}[M^{2}(P+1)+M(6P+2)+P+1]+M^{2}N}$  \\[4pt]
            %\hline
            ESPRIT \cite{Roy} & $\mathit{2M^{2}P+M(P^{2}-2P+8N^{2})+8P^{3}-P^{2} }$  \\[4pt]
            \\
            & $\mathit{\tau[3M^{3}+M^{2}(3P+2)+M(\frac{5}{2}P^{2}-\frac{3}{2}P+8N^{2})}$  \\[2pt]

            &$\mathit{+P^{2}(\frac{17}{2}P+\frac{1}{2})+1]}$ \\[1pt]

            TS-ESPRIT \cite{Pinto}& $\mathit{+[2M^{3}+M^{2}(3P)+M(\frac{5}{2}P^{2}-\frac{3}{2}P+8N^{2})}$ \\[2pt]

            & $\mathit{+P^{2}(\frac{17}{2}P+\frac{1}{2})]}$\\[3pt]

            \hline

    \end{tabular}}

    \label{Comput_Complexity1}
\end{table}
%
%\begin{table}[!h]
%    \caption{Computational complexity - other algorithms}
%    \vspace{2mm}
%    \centering
%    \scalebox{0.92}{
%    \begin{tabular}{|l|l|p{6cm}|}
%        \hline
%        Algorithm & Multiplications \\[2pt]
%        \hline
%        MUSIC \cite{schimdt} & $\mathit{\frac{180}{\Delta}[M^{2}+M(2-P)-P]+8MN^{2}}$  \\[2pt]
%        \hline
%        root-MUSIC \cite{Barabell} & $\mathit{2M^{3}-M^{2}P+8MN^{2}}$  \\[2pt]
%        \hline
%        AVF \cite{Grover} & $\mathit{\frac{180}{\Delta}[M^{2}(3P+1)+M(4P-2)+P+2]}$\\[1pt]
%        &$\mathit{+M^{2}N}$  \\[2pt]
%        \hline
%        CG \cite{Semira} & $\mathit{\frac{180}{\Delta}[M^{2}(P+1)+M(6P+2)+P+1]+M^{2}N}$  \\[2pt]
%        \hline
%        ESPRIT \cite{Roy} & $\mathit{2M^{2}P+M(P^{2}-2P+8N^{2})+8P^{3}-P^{2} }$  \\[2pt]
%        \hline
%        & $\mathit{\tau[3M^{3}+M^{2}(3P+2)+M(\frac{5}{2}P^{2}-\frac{3}{2}P+8N^{2})}$  \\[1pt]
%
%        &$\mathit{+P^{2}(\frac{17}{2}P+\frac{1}{2})+1]}$ \\[1pt]
%
%        TS-ESPRIT \cite{Pinto}& $\mathit{+[2M^{3}+M^{2}(3P)+M(\frac{5}{2}P^{2}-\frac{3}{2}P+8N^{2})}$ \\[1pt]
%
%        & $\mathit{+P^{2}(\frac{17}{2}P+\frac{1}{2})]}$\\[2pt]
%
%        \hline
%    \end{tabular}}
%
%    \label{Comput_Complexity2}
%\end{table}
%
\section{Simulations}
\label{simulations}

In this section, we evaluate the performance of the proposed MS-KAI
CG-FB and MS-KAI-CG algorithms, the ordinary CG \cite{Semira} and
the forward-backward spatially smoothed CG (CG-FB)
\cite{Semira,Pillai}, the ESPRIT \cite{Roy}, and the MUSIC
\cite{schimdt} algorithms in terms of the root mean-square error
(RMSE) and the probability of resolution (PR). The RMSE and the
RMSE(dB) are defined  as
\begin{equation}
\centering \mathrm{RMSE}
=\sqrt{\frac{1}{S\:P}\sum\limits^{S}_{s=1}\sum\limits^{P}_{p=1}\bm(\theta_p
    -\bm \hat{\theta}_p(s))^{2}}, \label{RMSE_run}
\end{equation}
and
\begin{align}
\mathrm{RMSE\left( dB\right) }=
10 \log_{10}\left(\frac{RMSE^{o}}{1^{o}} \right),
\label{RMSE_dB}
\end{align}
{Since there were large gaps between some of the curves which form
the figure for assessing the RMSE performance  of the MS-KAI-CG-FB
in terms of degrees, we have  plotted the curves in dB in order to
compact them and to compare them to the square root of the
deterministic CRB \cite{Stoica4}.} To assess the performance in
terms of PR, we take into account the criterion of \cite{Stoica3},
in which two sources with DoA $\theta_{1}$ and $\theta_{2}$  are
said to be resolved if their respective estimates $\hat{\theta}_{1}$
and $\hat{\theta}_{2}$ are such that both $\left|\hat{\theta}_{1}
-\theta_{1}\right|$ and $\left|\hat{\theta}_{2} -\theta_{2}\right|$
are less than $\left|\theta_{1} -\theta_{2}\right|/2$. We have set
the search step to $ \varDelta=0.2^{o} $ in all algorithms that make
use of peak search.

\begin{figure}[!h]
    \centering % para centralizarmos a figura
    \includegraphics[width=8.4cm,height=6.3cm]{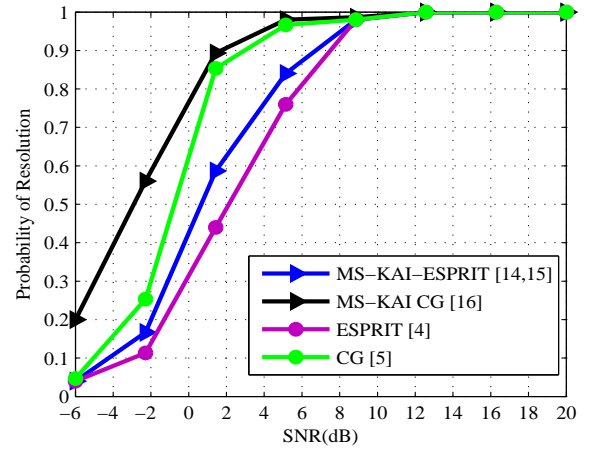} % leia abaixo
    %\vspace{-1.0em}
    \caption{\textit{Probability of resolution versus SNR with} $P=2$, $M=12$, $N=100$, $L=150$ \textit{runs}, $\xi\left(\theta \right) =2.0^{o}.$}
    \label{figura:PR_12_100_step02_2ponto0_graus_150_uncorrel}
\end{figure}

\begin{figure}[!h]
    \centering % para centralizarmos a figura
    \includegraphics[width=8.4cm, height=6.3cm]{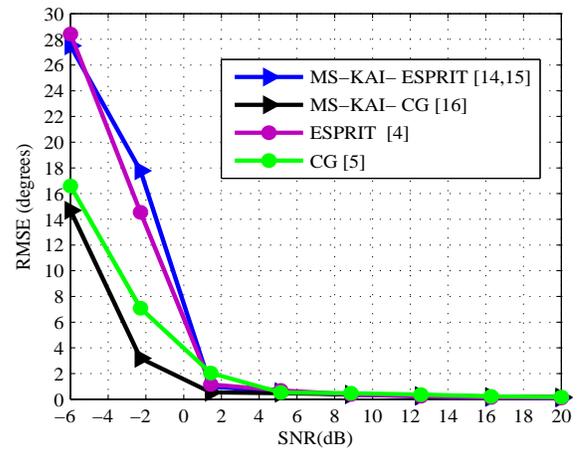} % leia abaixo
    %\vspace{-1.0em}
    \caption{\textit{RMSE in degrees versus SNR with} $P=2$, $M=12$, $N=100$, $L=150$ \textit{runs}, $\xi\left(\theta \right) =2.0^{o}$.}
    \label{figura:RMSE_12_100_step02_2ponto0_graus_150_uncorrel}
\end{figure}

We first consider a scenario with $\mathit{P=2}$ uncorrelated
complex Gaussian signals with equal power impinging on a ULA with
$\mathit{N=12}$ sensors. The sources have been separated by
$\xi\left(\theta \right) =2.0^{o}$, at $\mathrm(15^{o},17^{o})$, and
the number of available snapshots was set to $\mathit{N}=100$. The
computations of RMSE  have used 150 independent trials. In Fig.
\ref{figura:PR_12_100_step02_2ponto0_graus_150_uncorrel}, we show
the PR against the SNR, whereas in Fig.
\ref{figura:RMSE_12_100_step02_2ponto0_graus_150_uncorrel} the RMSE
performance against the SNR is depicted. From the curves it can be
noticed the improvement of the performance of MS-KAI-CG in terms of
both PR and RMSE as a result of the improved estimate of the data
covariance matrix.

\begin{figure}[!h]
    \centering % para centralizarmos a figura
    \includegraphics[width=8.4cm, height=6.3cm]{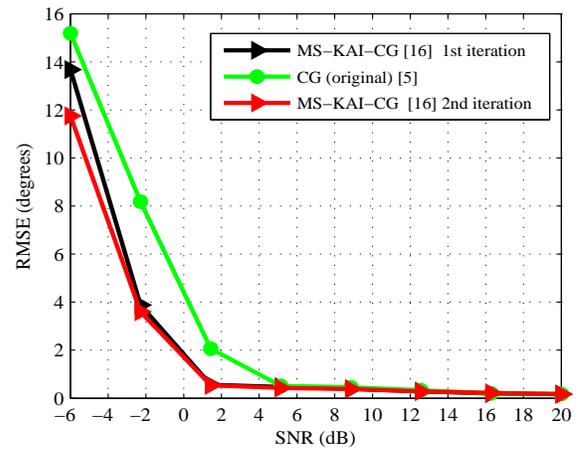} % leia abaixo
    %\vspace{-1.0em}
    \caption{\textit{Influence of the iterations in terms of RMSE in degrees  versus SNR with} $P=2$, $M=12$, $N=100$, $L=150$ \textit{runs}, $\xi\left(\theta \right) =2.0^{o}$.}
    \label{figura:RMSE_influ_iter_12_100_step02_2ponto0_150_uncorrel}
\end{figure}

In Fig.
\ref{figura:RMSE_influ_iter_12_100_step02_2ponto0_150_uncorrel}, we
show the influence of the iterations carried out at the second step.
It can be noticed the gradual and consistent enhancement of the
performance of MS-KAI-CG in terms of RMSE as a result of the
increasing number of iterations.

\begin{figure}[!h]
    \centering % para centralizarmos a figura
    \includegraphics[width=8.4cm, height=6.3cm]{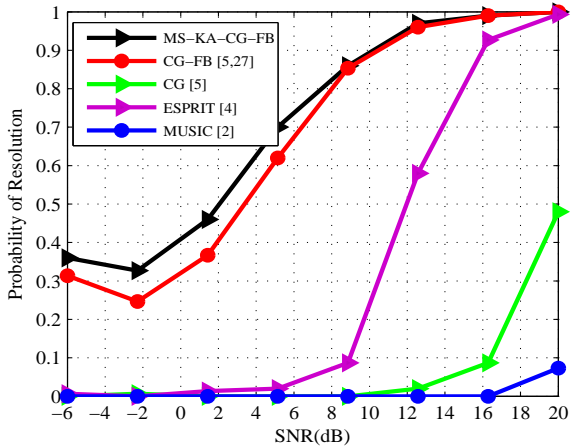} % leia abaixo
    %\vspace{-1.0em}
    \caption{\textit{ Probability of resolution versus SNR with} $P=2$, $M=12$, $N=70$, $L=150$ \textit{runs}, $\xi\left(\theta \right) =2.0^{o}$.}
    \label{figura:PR_MS_KAI_FB_CG_2fon_9_1_2gr_12sen_70snap_150runs}
\end{figure}

\begin{figure}[!h]
    \centering % para centralizarmos a figura
    \includegraphics[width=8.4cm, height=6.3cm]{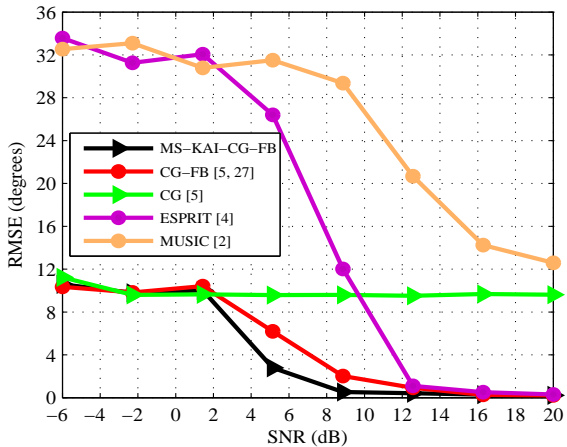} % leia abaixo
    %\vspace{-1.0em}
    \caption{\textit{RMSE in degrees versus SNR with} $P=2$, $M=12$, $N=70$, $L=150$ \textit{runs}, $\xi\left(\theta \right) =2.0^{o}$.}
    \label{figura:RMSE_MS_KAI_FB_CG_2fon_9_1_2gr_12sen_70snap_150runs}
\end{figure}

\begin{figure}[!h]
    \centering % para centralizarmos a figura
    \includegraphics[width=8.4cm, height=6.3cm]{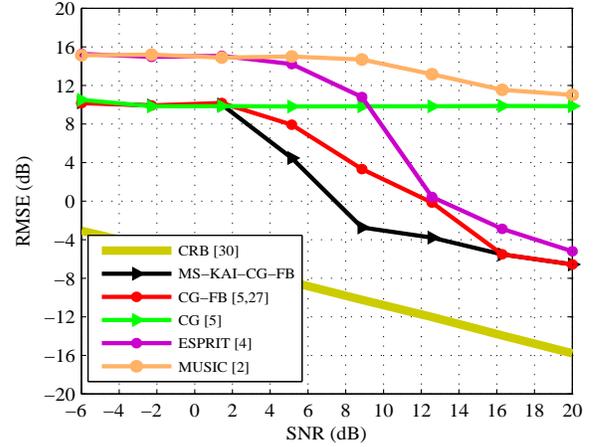} % leia abaixo
    %\vspace{-1.0em}
    \caption{\textit{ RMSE and the square root of CRB in dB versus SNR with} $P=2$, $M=12$, $N=70$, $L=150$ \textit{runs}, $\xi\left(\theta \right) =2.0^{o}$.}
    \label{figura:RMSE_dB_MS_KAI_FB_CG_2fon_9_1_2gr_12sen_70snap_150runs}
\end{figure}

In the next examples, we have studied the performance of the
proposed MS-KAI-CG-FB in the presence of strongly correlated closely
spaced sources. To this end, we consider a scenario composed of
Gaussian signals with equal power impinging on a ULA. In particular,
we have $P=2$ sources separated by $\xi\left(\theta \right)
=2.0^{o}$, at $\left(15^{o},17^{o} \right) $,  $M=12$ sensors and
$N=70$ snapshots. We have employed $L=150 $ trials for these
simulations. The source signals are correlated according
to the following correlation matrix: % $\bm R_{ss}$
\begin{equation}
\bm R_{ss}=\sigma_{s}^{2}\begin{bmatrix}
\label{signals_correlated_matrix1}
1   &  0.9    \\
0.9 &   1    \\
\end{bmatrix}.
\end{equation}
In Fig.
\ref{figura:PR_MS_KAI_FB_CG_2fon_9_1_2gr_12sen_70snap_150runs}, we
can notice that in terms of PR the suggested MS-KAI-CG-FB outperforms
the ordinary CG algorithm equipped with forward-backward spatial
smoothing, denoted as CG-FB, the ordinary CG algorithm, MUSIC and
ESPRIT in most of the considered range of SNR values.
In Fig.
\ref{figura:RMSE_MS_KAI_FB_CG_2fon_9_1_2gr_12sen_70snap_150runs}, we
can see that in terms of RMSE  the proposed MS-KAI-CG-FB provides
the best  performance in the range $ \left[1.8\;\:16\right]$ dB. It
can also be seen that in the ranges $ \left[-6\;\:1.8\right)$ dB and
$ \left(16\;\:20\right]$ dB its performance is similar to the best.
{This performance can be better noticed in Fig.
\ref{figura:RMSE_dB_MS_KAI_FB_CG_2fon_9_1_2gr_12sen_70snap_150runs},
which shows  the RMSE of all  curves which form the preceding
graphic and the square root of the deterministic CRB \cite{Stoica4},
all of them in terms of dB.}

We note that other transmit and receive processing structures can
also be considered following the approaches reported in
\cite{sintprec,sintprec2,lcrbd,gbd,wlbd,mbthp,rmbthp,bbprec} and
\cite{stbcmimo,jio,jidf,smtvb,jiomimo,rrmber,mberdf,spa,mfsic,mfdf,mbdf,did,tds,armo,badstc,1bitidd,baplnc,bfpeg,emd}.

\section{Conclusions}
We have developed the MS-KAI-CG algorithm and its version equipped
with forward-backward spatial smoothing termed MS-KAI-CG-FB
algorithm. Both approaches exploit the knowledge of signals and the
structure of the data covariance matrix, which are acquired on line
and used to subtract unwanted terms, thereby improving the
performance of existing CG-based DoA estimation algorithms. In
scenarios composed of two uncorrelated closely-spaced sources and a
sufficient number of snapshots, the MS-KAI-CG algorithm has shown
its superiority in terms of probability of resolution and RMSE over
existing algorithms, including the original CG, in the low and
medium levels of SNR, i.e., in the range $\left[ -6\;\:4\right]dB $.
In scenarios in which the uncorrelated signals previously considered
were replaced with strongly correlated signals, the comparisons of
MS-KAI-CG-FB algorithm with existing algorithms, including the
original CG and its version equipped with forward-backward spatial
smoothing (CG-FB), have shown a superior accuracy to the
MS-KAI-CG-FB algorithm in the following significant ranges: in
$\left[-6\;\: 8\right]dB$, for probability of resolution; and
$\left[2\;\: 16\right] dB$, for RMSE. Based on the significant
improvements pointed out, we can consider that MS-KAI-CG and
MS-KAI-CG-FB for dealing with correlated sources have excellent
potential for applications with significant data records in
large-scale sensor array systems for wireless communications, radar
and other applications with large sensor arrays.


\begin{thebibliography}{}
%
\bibitem{Vantrees}
Van Trees, H. L.: 'Optimum Array Processing',  (Wiley,New York, 2002).

\bibitem{locsme}
H. Ruan and R. C. de Lamare, ``Robust Adaptive Beamforming Using a
Low-Complexity Shrinkage-Based Mismatch Estimation Algorithm,"
\emph{IEEE Sig. Proc. Letters.}, Vol. 21, No. 1, pp 60-64, 2014.

\bibitem{elnashar}
A. Elnashar, ``Efficient implementation of robust adaptive
beamforming based on worst-case performance optimization," IET
Signal Process., Vol. 2, No. 4, pp. 381-393, Dec 2008.

\bibitem{manikas}
J. Zhuang and A. Manikas, ``Interference cancellation beamforming
robust to pointing errors," IET Signal Process., Vol. 7, No. 2, pp.
120-127, April 2013.

\bibitem{cgbf}
L. Wang and R. C. de Lamare, ``Constrained adaptive filtering
algorithms based on conjugate gradient techniques for beamforming,"
IET Signal Process., Vol. 4, No. 6, pp. 686-697, Feb 2010.

\bibitem{okspme}
H. Ruan and R. C. de Lamare, "Robust Adaptive Beamforming Based on
Low-Rank and Cross-Correlation Techniques," in IEEE Transactions on
Signal Processing, vol. 64, no. 15, pp. 3919-3932, 1 Aug.1, 2016.

\bibitem{r19}
H. Ruan and R. C. de Lamare, ``Low-Complexity Robust Adaptive
Beamforming Based on Shrinkage and Cross-Correlation," \emph{19th
International ITG Workshop on Smart Antennas}, pp 1-5, March 2015.

\bibitem{scharf}
L. L. Scharf and D. W. Tufts, ``Rank reduction for modeling
stationary signals," \textit{IEEE Transactions on Acoustics, Speech
and Signal Processing}, vol. ASSP-35, pp. 350-355, March 1987.

\bibitem{bar-ness} A. M. Haimovich
and Y. Bar-Ness, ``An eigenanalysis interference canceler," {\it
IEEE Trans. on Signal Processing}, vol. 39, pp. 76-84, Jan. 1991.

\bibitem{pados99} D. A. Pados and S. N. Batalama "Joint space-time
auxiliary vector filtering for DS/CDMA systems with antenna arrays"
\textit{ IEEE Transactions on Communications}, vol. 47, no. 9, pp.
1406 - 1415, 1999.



\bibitem{reed98} J. S. Goldstein, I. S. Reed and L. L. Scharf
"A multistage representation of the Wiener filter based on
orthogonal projections" \textit{IEEE Transactions on Information
Theory}, vol. 44, no. 7, 1998.

\bibitem{hua}
Y. Hua, M. Nikpour and P. Stoica, "Optimal reduced rank estimation
and filtering," IEEE Transactions on Signal Processing, pp. 457-469,
Vol. 49, No. 3, March 2001.


\bibitem{goldstein}
M. L. Honig and J. S. Goldstein, ``Adaptive reduced-rank
interference suppression based on the multistage Wiener filter,"
\textit{ IEEE Transactions on Communications}, vol. 50, no. 6, June
2002.

\bibitem{santos}
E. L. Santos and M. D. Zoltowski, ``On Low Rank MVDR Beamforming
using the Conjugate Gradient Algorithm", \textit{Proc. IEEE
International Conference on Acoustics, Speech and Signal
Processing}, 2004.

\bibitem{qian}
Q. Haoli and S.N. Batalama, ``Data record-based criteria for the
selection of an auxiliary vector estimator of the MMSE/MVDR filter",
\textit{IEEE Transactions on Communications}, vol. 51, no. 10, Oct.
2003, pp. 1700 - 1708.

\bibitem{delamarespl07}
R. C. de Lamare and R. Sampaio-Neto, ``Reduced-Rank Adaptive
Filtering Based on Joint Iterative Optimization of Adaptive
Filters", \textit{IEEE Signal Processing Letters}, Vol. 14, no. 12,
December 2007.

\bibitem{xutsa}
Z. Xu and M.K. Tsatsanis, ``Blind adaptive algorithms for minimum
variance CDMA receivers," \textit{IEEE Trans. Communications}, vol.
49, No. 1, January 2001.

\bibitem{delamaretsp}
R. C. de Lamare and R. Sampaio-Neto, ``Low-Complexity Variable
Step-Size Mechanisms for Stochastic Gradient Algorithms in Minimum
Variance CDMA Receivers", \textit{IEEE Trans. Signal Processing},
vol. 54, pp. 2302 - 2317, June 2006.

\bibitem{kwak}
C. Xu, G. Feng and K. S. Kwak, ``A Modified Constrained Constant
Modulus Approach to Blind Adaptive Multiuser Detection," \textit{
IEEE Trans. Communications}, vol. 49, No. 9, 2001.

\bibitem{xu&liu}
Z. Xu and P. Liu, ``Code-Constrained Blind Detection of CDMA Signals
in Multipath Channels," \textit{ IEEE Sig. Proc. Letters}, vol. 9,
No. 12, December 2002.

%\bibitem{BEACON_JIO_Clarke}
%P. Clarke and R. C. de Lamare,
%"Set-Membership Reduced-Rank BEACON Algorithm based on Joint Iterative Optimization of Adaptive Filters", \textit{IEEE International Symposium on Circuits and Systems}, Paris, June, 2010.
%
\bibitem{delamareccm}
R. C. de Lamare and R. Sampaio Neto, "Blind Adaptive
Code-Constrained Constant Modulus Algorithms for CDMA Interference
Suppression in Multipath Channels", \textit{ IEEE Communications
Letters}, vol 9. no. 4, April, 2005.

\bibitem{wcccm}
L. Landau, R. C. de Lamare and M. Haardt, ``Robust adaptive
beamforming algorithms using the constrained constant modulus
criterion," IET Signal Processing, vol.8, no.5, pp.447-457, July
2014.

\bibitem{delamareelb}
R. C. de Lamare, ``Adaptive Reduced-Rank LCMV Beamforming Algorithms
Based on Joint Iterative Optimisation of Filters",
\textit{Electronics Letters}, vol. 44, no. 9, 2008.


\bibitem{jidf}
R. C. de Lamare and R. Sampaio-Neto, ``Adaptive Reduced-Rank
Processing Based on Joint and Iterative Interpolation, Decimation
and Filtering", \textit{IEEE Transactions on Signal Processing},
vol. 57, no. 7, July 2009, pp. 2503 - 2514.

\bibitem{delamarecl}
R. C. de Lamare and Raimundo Sampaio-Neto, ``Reduced-rank
Interference Suppression for DS-CDMA based on Interpolated FIR
Filters", \textit{IEEE Communications Letters}, vol. 9, no. 3, March
2005.

\bibitem{delamaresp}
R. C. de Lamare and R. Sampaio-Neto, ``Adaptive Reduced-Rank MMSE
Filtering with Interpolated FIR Filters and Adaptive Interpolators",
\textit{IEEE Signal Processing Letters}, vol. 12, no. 3, March,
2005.

\bibitem{delamaretvt}
R. C. de Lamare and R. Sampaio-Neto, ``Adaptive Interference
Suppression for DS-CDMA Systems based on Interpolated FIR Filters
with Adaptive Interpolators in Multipath Channels", \textit{IEEE
Trans. Vehicular Technology}, Vol. 56, no. 6, September 2007.

\bibitem{jioel}
R. C. de Lamare, ``Adaptive Reduced-Rank LCMV Beamforming Algorithms
Based on Joint Iterative Optimisation of Filters," Electronics
Letters, 2008.


\bibitem{delamarespl07}
R. C. de Lamare and R. Sampaio-Neto, ``Reduced-rank adaptive
filtering based on joint iterative optimization of adaptive
filters",  \textit{IEEE Signal Process. Lett.}, vol. 14, no. 12, pp.
980-983, Dec. 2007.

\bibitem{delamare_ccmmswf}
R. C. de Lamare, M. Haardt, and R. Sampaio-Neto, ``Blind Adaptive
Constrained Reduced-Rank Parameter Estimation based on Constant
Modulus Design for CDMA Interference Suppression", \textit{IEEE
Transactions on Signal Processing}, June 2008.

\bibitem{jidf_echo}
M. Yukawa, R. C. de Lamare and R. Sampaio-Neto, ``Efficient Acoustic
Echo Cancellation With Reduced-Rank Adaptive Filtering Based on
Selective Decimation and Adaptive Interpolation," IEEE Transactions
on Audio, Speech, and Language Processing, vol.16, no. 4, pp.
696-710, May 2008.

\bibitem{delamaretvt10}
R. C. de Lamare and R. Sampaio-Neto, ``Reduced-rank space-time
adaptive interference suppression with joint iterative least squares
algorithms for spread-spectrum systems," \textit{IEEE Trans. Vehi.
Technol.}, vol. 59, no. 3, pp. 1217-1228, Mar. 2010.

\bibitem{delamaretvt2011ST}
R. C. de Lamare and R. Sampaio-Neto, ``Adaptive reduced-rank
equalization algorithms based on alternating optimization design
techniques for MIMO systems," \textit{IEEE Trans. Vehi. Technol.},
vol. 60, no. 6, pp. 2482-2494, Jul. 2011.


\bibitem{delamare10}
R. C. de Lamare, L. Wang, and R. Fa, ``Adaptive reduced-rank LCMV
beamforming algorithms based on joint iterative optimization of
filters: Design and analysis," Signal Processing, vol. 90, no. 2,
pp. 640-652, Feb. 2010.

\bibitem{fa10}
R. Fa, R. C. de Lamare, and L. Wang, ``Reduced-Rank STAP Schemes for
Airborne Radar Based on Switched Joint Interpolation, Decimation and
Filtering Algorithm," \textit{IEEE Transactions on Signal
Processing}, vol.58, no.8, Aug. 2010, pp.4182-4194.

\bibitem{lei09}
L. Wang and R. C. de Lamare, "Low-Complexity Adaptive Step Size
Constrained Constant Modulus SG Algorithms for Blind Adaptive
Beamforming", \textit{Signal Processing}, vol. 89, no. 12, December
2009, pp. 2503-2513.

\bibitem{ccmavf}
L. Wang and R. C. de Lamare, ``Adaptive Constrained Constant Modulus
Algorithm Based on Auxiliary Vector Filtering for Beamforming," IEEE
Transactions on Signal Processing, vol. 58, no. 10, pp. 5408-5413,
Oct. 2010.


\bibitem{lei10}
L. Wang, R. C. de Lamare, M. Yukawa, "Adaptive Reduced-Rank
Constrained Constant Modulus Algorithms Based on Joint Iterative
Optimization of Filters for Beamforming," \textit{IEEE Transactions
on Signal Processing}, vol.58, no.6, June 2010, pp.2983-2997.

\bibitem{jio_ccm}
L. Wang, R. C. de Lamare and M. Yukawa, ``Adaptive reduced-rank
constrained constant modulus algorithms based on joint iterative
optimization of filters for beamforming", IEEE Transactions on
Signal Processing, vol.58, no. 6, pp. 2983-2997, June 2010.

\bibitem{ccmavf}
L. Wang and R. C. de Lamare, ``Adaptive constrained constant modulus
algorithm based on auxiliary vector filtering for beamforming", IEEE
Transactions on Signal Processing, vol. 58, no. 10, pp. 5408-5413,
October 2010.

\bibitem{stap_jio}
R. Fa and R. C. de Lamare, ``Reduced-Rank STAP Algorithms using
Joint Iterative Optimization of Filters," IEEE Transactions on
Aerospace and Electronic Systems, vol.47, no.3, pp.1668-1684, July
2011.

\bibitem{zhaocheng}
Z. Yang, R. C. de Lamare and X. Li, ``L1-Regularized STAP Algorithms
With a Generalized Sidelobe Canceler Architecture for Airborne
Radar," IEEE Transactions on Signal Processing, vol.60, no.2,
pp.674-686, Feb. 2012.

\bibitem{zhaocheng2}
Z. Yang, R. C. de Lamare and X. Li, ``Sparsity-aware space–time
adaptive processing algorithms with L1-norm regularisation for
airborne radar", IET signal processing, vol. 6, no. 5, pp. 413-423,
2012.

\bibitem{arh_eusipco}
Neto, F.G.A.; Nascimento, V.H.; Zakharov, Y.V.; de Lamare, R.C.,
"Adaptive re-weighting homotopy for sparse beamforming," in Signal
Processing Conference (EUSIPCO), 2014 Proceedings of the 22nd
European , vol., no., pp.1287-1291, 1-5 Sept. 2014

\bibitem{arh_taes}
Almeida Neto, F.G.; de Lamare, R.C.; Nascimento, V.H.; Zakharov,
Y.V.,``Adaptive reweighting homotopy algorithms applied to
beamforming," IEEE Transactions on Aerospace and Electronic Systems,
vol.51, no.3, pp.1902-1915, July 2015.

\bibitem{dfjio}
L. Wang, R. C. de Lamare and M. Haardt, ``Direction finding
algorithms based on joint iterative subspace optimization," IEEE
Transactions on Aerospace and Electronic Systems, vol.50, no.4,
pp.2541-2553, October 2014.

\bibitem{rdrab}
S. D. Somasundaram, N. H. Parsons, P. Li and R. C. de Lamare,
``Reduced-dimension robust capon beamforming using Krylov-subspace
techniques," IEEE Transactions on Aerospace and Electronic Systems,
vol.51, no.1, pp.270-289, January 2015.

\bibitem{dcg_conf}
S. Xu and R.C de Lamare, , \textit{Distributed conjugate gradient
strategies for distributed estimation over sensor networks}, Sensor
Signal Processing for Defense SSPD, September 2012.


\bibitem{dcg}
S. Xu, R. C. de Lamare, H. V. Poor, ``Distributed Estimation Over
Sensor Networks Based on Distributed Conjugate Gradient Strategies",
IET Signal Processing, 2016 (to appear).

\bibitem{dce}
S. Xu, R. C. de Lamare and H. V. Poor, \textit{Distributed
Compressed Estimation Based on Compressive Sensing}, IEEE Signal
Processing letters, vol. 22, no. 9, September 2014.

\bibitem{drr_conf}
S. Xu, R. C. de Lamare and H. V. Poor, ``Distributed reduced-rank
estimation based on joint iterative optimization in sensor
networks," in Proceedings of the 22nd European Signal Processing
Conference (EUSIPCO), pp.2360-2364, 1-5, Sept. 2014

\bibitem{dta_conf1}
S. Xu, R. C. de Lamare and H. V. Poor, ``Adaptive link selection
strategies for distributed estimation in diffusion wireless
networks," in Proc. IEEE International Conference onAcoustics,
Speech and Signal Processing (ICASSP),  , vol., no., pp.5402-5405,
26-31 May 2013.

\bibitem{dta_conf2}
S. Xu, R. C. de Lamare and H. V. Poor, ``Dynamic topology adaptation
for distributed estimation in smart grids," in Computational
Advances in Multi-Sensor Adaptive Processing (CAMSAP), 2013 IEEE 5th
International Workshop on , vol., no., pp.420-423, 15-18 Dec. 2013.

\bibitem{dta_ls}
S. Xu, R. C. de Lamare and H. V. Poor, ``Adaptive Link Selection
Algorithms for Distributed Estimation", EURASIP Journal on Advances
in Signal Processing, 2015.

\bibitem{song}
N. Song, R. C. de Lamare, M. Haardt, and M. Wolf, ``Adaptive Widely
Linear Reduced-Rank Interference Suppression based on the
Multi-Stage Wiener Filter," IEEE Transactions on Signal Processing,
vol. 60, no. 8, 2012.

\bibitem{wljio}
N. Song, W. U. Alokozai, R. C. de Lamare and M. Haardt, ``Adaptive
Widely Linear Reduced-Rank Beamforming Based on Joint Iterative
Optimization,"  IEEE Signal Processing Letters, vol.21, no.3, pp.
265-269, March 2014.

\bibitem{barc}
R.C. de Lamare, R. Sampaio-Neto and M. Haardt, "Blind Adaptive
Constrained Constant-Modulus Reduced-Rank Interference Suppression
Algorithms Based on Interpolation and Switched Decimation,"
\textit{IEEE Trans. on Signal Processing},  vol.59, no.2,
pp.681-695, Feb. 2011.

\bibitem{jiomber}
Y. Cai, R. C. de Lamare, ``Adaptive Linear Minimum BER Reduced-Rank
Interference Suppression Algorithms Based on Joint and Iterative
Optimization of Filters," IEEE Communications Letters, vol.17, no.4,
pp.633-636, April 2013.

\bibitem{saalt}
R. C. de Lamare and R. Sampaio-Neto, ``Sparsity-Aware Adaptive
Algorithms Based on Alternating Optimization and Shrinkage," IEEE
Signal Processing Letters, vol.21, no.2, pp.225,229, Feb. 2014.

\bibitem{mmimo}
R. C. de Lamare, ``Massive MIMO Systems: Signal Processing
Challenges and Future Trends", Radio Science Bulletin, December
2013.

\bibitem{wence}
W. Zhang, H. Ren, C. Pan, M. Chen, R. C. de Lamare, B. Du and J.
Dai, ``Large-Scale Antenna Systems With UL/DL Hardware Mismatch:
Achievable Rates Analysis and Calibration", IEEE Trans. Commun.,
vol.63, no.4, pp. 1216-1229, April 2015.

\bibitem{spa}
R. C. De Lamare and R. Sampaio-Neto, "Minimum Mean-Squared Error
Iterative Successive Parallel Arbitrated Decision Feedback Detectors
for DS-CDMA Systems," in IEEE Transactions on Communications, vol.
56, no. 5, pp. 778-789, May 2008.

\bibitem{mbdf}
R. C. de Lamare, "Adaptive and Iterative Multi-Branch MMSE Decision
Feedback Detection Algorithms for Multi-Antenna Systems," in IEEE
Transactions on Wireless Communications, vol. 12, no. 10, pp.
5294-5308, October 2013.

\bibitem{rrmber}
Y. Cai, R. C. de Lamare, B. Champagne, B. Qin and M. Zhao, "Adaptive
Reduced-Rank Receive Processing Based on Minimum Symbol-Error-Rate
Criterion for Large-Scale Multiple-Antenna Systems," in IEEE
Transactions on Communications, vol. 63, no. 11, pp. 4185-4201, Nov.
2015.

\bibitem{bfidd}
A. G. D. Uchoa, C. T. Healy and R. C. de Lamare, "Iterative
Detection and Decoding Algorithms for MIMO Systems in Block-Fading
Channels Using LDPC Codes," in IEEE Transactions on Vehicular
Technology, vol. 65, no. 4, pp. 2735-2741, April 2016.

\bibitem{did}
P. Li and R. C. de Lamare, "Distributed Iterative Detection With
Reduced Message Passing for Networked MIMO Cellular Systems," in
IEEE Transactions on Vehicular Technology, vol. 63, no. 6, pp.
2947-2954, July 2014.

\bibitem{mbthp}
K. Zu, R. C. de Lamare and M. Haardt, "Multi-Branch
Tomlinson-Harashima Precoding Design for MU-MIMO Systems: Theory and
Algorithms," in IEEE Transactions on Communications, vol. 62, no. 3,
pp. 939-951, March 2014.

\bibitem{wlbd}
W. Zhang et al., "Widely Linear Precoding for Large-Scale MIMO with
IQI: Algorithms and Performance Analysis," in IEEE Transactions on
Wireless Communications, vol. 16, no. 5, pp. 3298-3312, May 2017.

\bibitem{baplnc}
J. Gu, R. C. de Lamare and M. Huemer, "Buffer-Aided Physical-Layer
Network Coding With Optimal Linear Code Designs for Cooperative
Networks," in IEEE Transactions on Communications, vol. 66, no. 6,
pp. 2560-2575, June 2018.

\bibitem{schimdt}
Schmidt, R.: 'Multiple emitter location and signal parameter estimation',
\textit{IEEE Trans on Antennas and Propagation}, \textbf{34}, (3), pp 276-280, 1986.

\bibitem{Barabell}
Barabell, A. J.: 'Improving the resolution performance of eigenstructure-based
direction-finding algorithms', Proc. Int. Conf. on Acoustic Speech and Signal Processing ICASSP, Boston, pp. 336-339, 1983.

\bibitem{Roy}
Roy, R.  and  Kailath, T: 'Estimation of signal parameters via
rotational invariance techniques', \textit{IEEE Trans. Acoust., Speech and Signal Processing}, \textbf{37},  pp 984-995, 1989.

\bibitem{smcg}
L. Wang and R. C. DeLamare, "Low-Complexity Constrained Adaptive
Reduced-Rank Beamforming Algorithms," in IEEE Transactions on
Aerospace and Electronic Systems, vol. 49, no. 4, pp. 2114-2128,
October 2013.

\bibitem{rdrcb}
S. D. Somasundaram, N. H. Parsons, P. Li and R. C. de Lamare,
"Reduced-dimension robust capon beamforming using Krylov-subspace
techniques," in IEEE Transactions on Aerospace and Electronic
Systems, vol. 51, no. 1, pp. 270-289, January 2015.

\bibitem{Semira}
Semira, H., Belkacemi, Marcos, S.: 'High-resolution source
localization algorithm based on the Conjugate Gradient',
\textit{EURASIP Journal on Advances in Signal Processing}, 2007.

\bibitem{Steinwandt}
Steinwandt, J.,Lamare, R., Haardt, M.: 'Beamspace direction
finding based on the conjugate gradient and the auxiliary vector
filtering algorithms', \textit{Signal Processing}, \textbf{93}, (4), pp. 641-651, 2013.

\bibitem{Wang}
Wang, L., Lamare, R., Haardt, M.: 'Direction finding
algorithms based on joint iterative subspace optimization',
\textit{IEEE Transactions on Aerospace and Electronic Systems},
\textbf{50}, (4), pp. 2541-2553, 2014.

\bibitem{Qiu}
Qiu, L.,  Lamare, R., Zhao, M.: 'Reduced-Rank DOA
Estimation Algorithms Based on Alternating Low-Rank Decomposition',
\textit{IEEE Signal Processing Letters}\textbf{23}, (5), pp.
565-569, May 2016.

\bibitem{Pal1}
{Pal, P., Vaidyanathan, P.P.: 'A Novel Approach to Array Processing
With Enhanced Degrees of Freedom',\textit{IEEE Transactions on
Signal Processing},  \textbf{58}, (8), pp. 4167-4181, 2010.}

\bibitem{Pal2}
{Pal, P., Vaidyanathan, P.P.:'Sparse sensing with co-prime samplers
and arrays', \textit{IEEE Transactions on Signal Processing},
\textbf{59}, (2), pp. 573-586, 2011.}

\bibitem{Gu1}
{ Zhou, C., Gu, Y., Zhang, Y.D., Shi, Z., Jin, T.,
Wu,X.:'Compressive sensing-based coprime array direction-of-arrival
estimation', \textit{IET Communications},  \textbf{11}, (11), pp.
1719-1724, 2017.}

\bibitem{Gu2}
{Shi, Z., Zhou, C., Gu, Y., Goodman, N.A., Qu, F.:'Source Estimation
using Coprime Array: A Sparse Reconstruction Perspective',
\textit{IEEE Sensors Journal},  \textbf{17}, (3), pp. 755-765,
2017.}

\bibitem{Pinto}
Pinto, S.,  Lamare, R: 'Two-Step Knowledge-aided Iterative ESPRIT
Algorithm', Proc. IEEE Twenty First ITG Workshop on Smart Antennas,
Berlin, Germany, pp. 1-5, March 2017.

\bibitem{Pinto1}
Pinto, S.,  Lamare, R.: 'Multi-Step Knowledge-Aided Iterative ESPRIT
for Direction Finding', Proc. IEEE 22nd International Conference on
Digital Signal Processing, London, UK, pp. 1-5, August 2017.

\bibitem{Pinto4}
{Pinto, S.,  Lamare, R: 'Multi-Step Knowledge-Aided Iterative
ESPRIT: Design and Analysis', \textit{IEEE Transactions on Aerospace
and Electronic  Systems}, (early access),  pp 1-1,
2018.}

\bibitem{Pinto2} Pinto, S.,  Lamare, R.: ' Multi-Step
Knowledge-Aided Iterative Conjugate Gradient for Direction Finding',
Proc. IEEE 22nd ITG Workshop on Smart Antennas, Bochum, Germany,
2018.

\bibitem{Vorobyov1}
Shaghaghi, M., Vorobyov, S.: ' Iterative root-MUSIC algorithm for
DOA estimation', Proc. IEEE 5th  International Workshop on
Computational Advances in Multisensor Adaptive Processing, 2013.

\bibitem{Vorobyov2}
Shaghaghi, M., Vorobyov, S.: 'Subspace leakage analysis and improved
DOA estimation with small sample size', \textit{IEEE Trans. Signal
Processing},\textbf{63}, (12), pp 3251-3265, 2015.

\bibitem{Pinto3}
Pinto, S.,  Lamare, R.: 'Knowledge-Aided Parameter Estimation Based
on Conjugate Gradient Algorithms', 35th  Brazilian Communications
and Signal Processing Symposium, Sao Pedro, SP, Brazil, pp.1-5,
2017.

\bibitem{Steinwandt2}
Steinwandt, J.,Lamare, R., Haardt, M.: 'Knowledge-aided
direction finding based on Unitary ESPRIT', Proc IEEE 45th Asilomar Conference on Signals, Systems and
Computers, pp. 613-617, 2011.

\bibitem{Stoica2}
Stoica, P., Zhu, X., Guerci, J.: 'On using a priori
knowledge in space-time adaptive processing', \textit{IEEE Transactions on Signal Processing}, pp. 2598-2602, 2008.

\bibitem{Rappaport}
Liberti Jr,J., Rappaport,T.: 'Smart antennas for Wireless Communications: IS-95 and Third Generation CDMA Applications', Prentice Hall, 1999.

\bibitem{Schell}
Schell,S, Gardner, W.:'High Resolution Direction Finding', Handbook of Statistics, Elsevier, 1993.

\bibitem{Rissanen}
Rissanen,J.: 'Modeling by the Shortest Data Description', Automatica, \textbf{14}, pp 465-471, 1978.

\bibitem{Stoica}
Stoica, P.,  Nehorai, A.: 'Performance study of conditional and
unconditional direction-of-arrival estimation', \textit{IEEE Trans.
Acoust., Speech, Signal Processing},  \textbf{38}, (10), pp.
1783-1795, 1990,

\bibitem{Stoica3}
Stoica, P., Gershman, A.: 'Maximum-likelihood {DOA}  estimation by
data-supported grid search', \textit{IEEE Signal Processing
Letters}, pp. 273-275, 1999.

\bibitem{Pillai}
Pillai, S., Kwon B.: 'Forward/Backward spatial smoothing techniques
for coherent signal identification', \textit{IEEE Trans. Acoustics,
Speech, and Signal Processing}, pp. 8-15, 1989.

\bibitem{Haardt}
Thakre, A., Haardt, M., Giridhar, K: 'Single Snapshot Spatial
Smoothing with Improved Effective Array Aperture', \textit{IEEE
Signal Processing Letters}, \textbf{6}, 2009.

\bibitem{Steinwandt3}
{Steinwandt, J., Lamare, R. C., Haardt, M.: “Beamspace direction
finding based on the conjugate gradient algorithm”, 2011,
International ITG Workshop on Smart Antennas, pp.1-5, Feb. 2011,
included in IEEE Xplore in  April 2011.}

\bibitem{Stoica4}
Stoica, P., Nehorai, A.: 'MUSIC, maximum Likelihood, and
{Cram\'{e}r-Rao} Bound', \textit{IEEE Transactions on Acoustics,
Speech and Signal Processing}, pp. 720- 741, 1989.

\bibitem{Grover}
Grover, R., Pados, D., Medley, M.: 'Subspace direction finding with
an auxiliary-vector basis', \textit{IEEE Transactions on Signal Processing},  pp.758-763, 2007.

\bibitem{golub}
Golub, G. , van Loan, C.: "Matrix Computations", Johns Hopkins
University Press, 3rd edition, 1996.

\bibitem{sintprec}
Y. Cai, R. C. d. Lamare and R. Fa, "Switched Interleaving Techniques
with Limited Feedback for Interference Mitigation in DS-CDMA
Systems," in IEEE Transactions on Communications, vol. 59, no. 7,
pp. 1946-1956, July 2011.

\bibitem{sintprec2}
Y. Cai, R. C. de Lamare and D. Le Ruyet, "Transmit Processing
Techniques Based on Switched Interleaving and Limited Feedback for
Interference Mitigation in Multiantenna MC-CDMA Systems," in IEEE
Transactions on Vehicular Technology, vol. 60, no. 4, pp. 1559-1570,
May 2011.

\bibitem{lcrbd}
K. Zu and R. C. d. Lamare, "Low-Complexity Lattice Reduction-Aided
Regularized Block Diagonalization for MU-MIMO Systems," in IEEE
Communications Letters, vol. 16, no. 6, pp. 925-928, June 2012.

\bibitem{gbd}
K. Zu, R. C. de Lamare and M. Haardt, "Generalized Design of
Low-Complexity Block Diagonalization Type Precoding Algorithms for
Multiuser MIMO Systems," in IEEE Transactions on Communications,
vol. 61, no. 10, pp. 4232-4242, October 2013.

\bibitem{wlbd}
W. Zhang et al., "Widely Linear Precoding for Large-Scale MIMO with
IQI: Algorithms and Performance Analysis," in IEEE Transactions on
Wireless Communications, vol. 16, no. 5, pp. 3298-3312, May 2017.

\bibitem{mbthp}
K. Zu, R. C. de Lamare and M. Haardt, "Multi-Branch
Tomlinson-Harashima Precoding Design for MU-MIMO Systems: Theory and
Algorithms," in IEEE Transactions on Communications, vol. 62, no. 3,
pp. 939-951, March 2014.

\bibitem{rmbthp}
L. Zhang, Y. Cai, R. C. de Lamare and M. Zhao, "Robust Multibranch
Tomlinson–Harashima Precoding Design in Amplify-and-Forward MIMO
Relay Systems," in IEEE Transactions on Communications, vol. 62, no.
10, pp. 3476-3490, Oct. 2014.

\bibitem{bbprec}
L. T. N. Landau and R. C. de Lamare, "Branch-and-Bound Precoding for
Multiuser MIMO Systems With 1-Bit Quantization," in IEEE Wireless
Communications Letters, vol. 6, no. 6, pp. 770-773, Dec. 2017.


\bibitem{jidf}
R. C. de Lamare and R. Sampaio-Neto, "Adaptive Reduced-Rank
Processing Based on Joint and Iterative Interpolation, Decimation,
and Filtering," in IEEE Transactions on Signal Processing, vol. 57,
no. 7, pp. 2503-2514, July 2009.

\bibitem{jio}
R. C. de Lamare and R. Sampaio-Neto, "Reduced-Rank Adaptive
Filtering Based on Joint Iterative Optimization of Adaptive
Filters," in IEEE Signal Processing Letters, vol. 14, no. 12, pp.
980-983, Dec. 2007.

\bibitem{jiomimo}
R. C. de Lamare and R. Sampaio-Neto, "Adaptive Reduced-Rank
Processing Based on Joint and Iterative Interpolation, Decimation,
and Filtering," in IEEE Transactions on Signal Processing, vol. 57,
no. 7, pp. 2503-2514, July 2009.

\bibitem{smtvb}
R. C. de Lamare and P. S. R. Diniz, "Set-Membership Adaptive
Algorithms Based on Time-Varying Error Bounds for CDMA Interference
Suppression," in IEEE Transactions on Vehicular Technology, vol. 58,
no. 2, pp. 644-654, Feb. 2009.

\bibitem{stbcmimo}
R. C. De Lamare and R. Sampaio-Neto, "Blind adaptive MIMO receivers
for space-time block-coded DS-CDMA systems in multipath channels
using the constant modulus criterion," in IEEE Transactions on
Communications, vol. 58, no. 1, pp. 21-27, January 2010.

\bibitem{rrmber}
Y. Cai, R. C. de Lamare, B. Champagne, B. Qin and M. Zhao, "Adaptive
Reduced-Rank Receive Processing Based on Minimum Symbol-Error-Rate
Criterion for Large-Scale Multiple-Antenna Systems," in IEEE
Transactions on Communications, vol. 63, no. 11, pp. 4185-4201, Nov.
2015.

\bibitem{mberdf}
R. C. de Lamare and R. Sampaio-Neto, "Adaptive MBER decision
feedback multiuser receivers in frequency selective fading
channels," in IEEE Communications Letters, vol. 7, no. 2, pp. 73-75,
Feb. 2003.

\bibitem{spa}
R. C. De Lamare and R. Sampaio-Neto, "Minimum Mean-Squared Error
Iterative Successive Parallel Arbitrated Decision Feedback Detectors
for DS-CDMA Systems," in IEEE Transactions on Communications, vol.
56, no. 5, pp. 778-789, May 2008.

\bibitem{mfsic}
P. Li, R. C. de Lamare and R. Fa, "Multiple Feedback Successive
Interference Cancellation Detection for Multiuser MIMO Systems," in
IEEE Transactions on Wireless Communications, vol. 10, no. 8, pp.
2434-2439, August 2011.

\bibitem{mfdf}
P. Li and R. C. De Lamare, "Adaptive Decision-Feedback Detection
With Constellation Constraints for MIMO Systems," in IEEE
Transactions on Vehicular Technology, vol. 61, no. 2, pp. 853-859,
Feb. 2012.

%\bibitem{mbdf}
%R. C. de Lamare, "Adaptive and Iterative Multi-Branch MMSE Decision
%Feedback Detection Algorithms for Multi-Antenna Systems," in IEEE
%Transactions on Wireless Communications, vol. 12, no. 10, pp.
%5294-5308, October 2013.
%
\bibitem{did}
P. Li and R. C. de Lamare, "Distributed Iterative Detection With
Reduced Message Passing for Networked MIMO Cellular Systems", IEEE
Transactions on Vehicular Technology, vol. 63, no. 6, pp. 2947-2954,
2014.

\bibitem{tds}
P. Clarke and R. C. de Lamare, "Transmit Diversity and Relay
Selection Algorithms for Multirelay Cooperative MIMO Systems," in
IEEE Transactions on Vehicular Technology, vol. 61, no. 3, pp.
1084-1098, March 2012.

\bibitem{armo}
T. Peng, R. C. de Lamare and A. Schmeink, "Adaptive Distributed
Space-Time Coding Based on Adjustable Code Matrices for Cooperative
MIMO Relaying Systems," in IEEE Transactions on Communications, vol.
61, no. 7, pp. 2692-2703, July 2013.

\bibitem{badstc}
T. Peng and R. C. de Lamare, "Adaptive Buffer-Aided Distributed
Space-Time Coding for Cooperative Wireless Networks," in IEEE
Transactions on Communications, vol. 64, no. 5, pp. 1888-1900, May
2016.

\bibitem{bfidd}
A. G. D. Uchoa, C. T. Healy and R. C. de Lamare, "Iterative
Detection and Decoding Algorithms for MIMO Systems in Block-Fading
Channels Using LDPC Codes," in IEEE Transactions on Vehicular
Technology, vol. 65, no. 4, pp. 2735-2741, April 2016.

\bibitem{1bitidd}
Z. Shao, R. C. de Lamare and L. T. N. Landau, "Iterative Detection
and Decoding for Large-Scale Multiple-Antenna Systems With 1-Bit
ADCs," in IEEE Wireless Communications Letters, vol. 7, no. 3, pp.
476-479, June 2018.

\bibitem{baplnc}
J. Gu, R. C. de Lamare and M. Huemer, "Buffer-Aided Physical-Layer
Network Coding With Optimal Linear Code Designs for Cooperative
Networks," in IEEE Transactions on Communications, vol. 66, no. 6,
pp. 2560-2575, June 2018.

\bibitem{bfpeg}
A. G. D. Uchoa, C. Healy, R. C. de Lamare and R. D. Souza, "Design
of LDPC Codes Based on Progressive Edge Growth Techniques for Block
Fading Channels," in IEEE Communications Letters, vol. 15, no. 11,
pp. 1221-1223, November 2011.

\bibitem{emd}
C. T. Healy and R. C. de Lamare, "Design of LDPC Codes Based on
Multipath EMD Strategies for Progressive Edge Growth," in IEEE
Transactions on Communications, vol. 64, no. 8, pp. 3208-3219, Aug.
2016.

\end{thebibliography}
\end{document}